\documentclass[12pt,preprint]{aastex}

\shorttitle{LSB edge-on galaxies}
\shortauthors{Bizyaev and Kajsin}

\begin{document}

\baselineskip 12pt

\title{The stellar disk thickness of LSB galaxies}
                                                  
\author{D.Bizyaev\altaffilmark{1,2,3}} 
\email{dmbiz@sai.msu.ru}
\and  \author{S.Kajsin\altaffilmark{4}}
\altaffiltext{1}{Physics Department, University of Texas at El Paso, TX 79968}
\altaffiltext{2}{Sternberg Astronomical Institute, Moscow, 119899, Russia}
\altaffiltext{3}{Isaac Newton Institute of Chili, Moscow Branch}
\altaffiltext{4}{Special Astrophysical Observatory of RAS, pos. 
Nizhnij Arkhyz, 357147,  Karachaevo-Cherkessia, Russia}
\email{skai@sao.ru}

\begin{abstract}
We present surface photometry results for a sample of eleven edge-on 
galaxies observed with the 6m telescope at the Special Astrophysical 
Observatory (Russia). The photometric scale length, scale height, and 
central surface brightness of the stellar disks of our sample galaxies are 
estimated. We show that four galaxies in our sample, which are visually 
referred as objects of the lowest surface brightness class in the Revised 
Flat Galaxies Catalog, have bona fide low surface brightness (LSB) disks. 
We find from the comparison of photometric scales that the stellar disks of 
LSB galaxies are thinner than those of high surface brightness (HSB) ones. 
There is a clear correlation
between the central surface brightness of the stellar disk and its
vertical to radial scale ratio. The masses of spherical subsystems
(dark halo + bulge) and the dark halo masses are obtained for the sample
galaxies based on the thickness of their stellar disks.
The LSB galaxies tend to harbor more massive
spherical subsystems than the HSB objects, whereas no 
systematic difference in the dark halo masses between LSB and HSB galaxies
is found. At the same time, the inferred mass-to-luminosity ratio for the 
LSB disks appears to be systematically higher than for HSB disks.
\end{abstract}


\section{Introduction}

Low surface brightness spiral galaxies (hereafter, LSB galaxies) have
been studied extensively in recent years. Their main distinctive feature 
from "regular", high surface brightness (HSB) galaxies, is roughly a two 
magnitude lower central surface brightness of their stellar disks. They 
are thought to harbor massive dark halos \citep{deBlok03}. The LSB rotation 
curves are shallower in their central parts \citep{McG01}, which points 
toward a large dark matter fraction.

By observing the thickness of the stellar disk in a galaxy, one can 
constrain the relative mass of the dark halo \citep{Z91}. Until recently, 
only few edge-on LSB galaxies have been explored in detail (e.g. UGC 7321
\citet{Matt00} and IC 5249 \citet{vdK01}). 

We conducted a study of a small 
uniform sample of LSB and HSB galaxies observed with the same instrument 
to compare their structural parameters.
Here we present the results of photometric observations
in the V and R bands of a sample of
eleven edge-on galaxies. The paper is structured as follows:
in section 2 we describe the sample of galaxies and
observations. In section 3 we discuss the data reduction
and present the structural parameters of our galaxies. In section
4 we use the inferred disk thickness to estimate the dark halo mass.
Section 5 contains a discussion of
selection effects and relations between the inferred parameters.
The main results are summarized in section 6.

\section{Sample of galaxies and observations}

Our sample is based on the Revised Catalog of Flat
Galaxies (\citet{RFGC}, RFGC hereafter).  All galaxies included in this
catalog are highly inclined objects. We select object from the
faintest surface brightness (SB) class (IV according to RFGC) as candidates
to LSB galaxies, and objects from intermediate or high surface
brightness classes as reference HSB objects. We narrowed the sample of
objects to galaxies large enough for srtuctural studies
(major axis size $>$2 \arcmin ~in RFGC) which fit inside the 
3.5 \arcmin ~field-of-view of our CCD imager. 
In three observing nights of our program we obtained data for 11 galaxies.

Photometric observations were performed with the prime focus
camera on the 6-m telescope at the Special Astrophysical Observatory
of the Russian Academy of Sciences. This setup provides a plate scale
of 0.2 arcsec/pixel and good sensitivity for faint regions of galaxies.
The data were taken on April 27, 28, and 30, 2000 in the Johnson-Cousins
V and R photometric bands. The V-band images were utilized mostly for
calibration purposes, while the R images were used for the measurements
of structural parameters. For most of the galaxies in our sample we made
two to four images shifted by a few pixels in both photometric bands.


The Landolt photometric standards \citep{Landolt92} were observed every
night. Table~1 summarizes our observations listing object names,
surface brightness class (according to RFGC), date of the observation,
total integration time in every photometric band, number of
exposures, and average seeing when the target was observed.

\section{Data reduction and results}

The data were reduced using standard tools in the 
MIDAS package. The images of galaxies and photometric standards were 
corrected for the bias and dark current, and flat fielded. 
The images were then sky subtracted, aligned, and
combined. We checked the quality of flat fielding and sky subtraction
by comparing the background level in those parts of the
image that are free of foreground stars and are located close to a sample
galaxy. The large-scale pattern of the background does not introduce 
uncertainties above 0.1\%. Three galaxies have very 
bright stars in their fields, which raises the large-scale background 
fluctuations up to 0.2\%.

Eight and 
twelve Landolt's stars from three selected areas were available on April 
27 and 30, respectively. The residuals for photometric solutions were 
0.$^m$02 for April 27 (in both V and R bands), and 0.$^m$04 (in both V
and R bands) for the night on April 30. The sky brightness level is 
tabulated in Table \ref{tab1}. The surface brightness corresponding 
to a 3 $\sigma$ level of the background noise in the final combined 
images is shown in Table 1 as well. 

The observing conditions were non-photometric during part
of the night on April 28. However, most of our galaxies have
the major axes photometric profiles in the R band published by 
\citet{IDK92}. It enables us to verify the calibration and to adjust 
it for the non-photometric night. The mean difference between 
the surface brightnesses we derived and those 
published by \citet{IDK92} is of the order of 0.$^m$3.
The largest source of the discrepance comes from the use of 
different procedures of the major axis profiles extraction.

Comparison of the sky brightness in R images 
can be used to estimate roughly the zero point of calibration for the 
objects taken on April 28. If we use this way of calibration, 
the R-band surface brightnesses of UGC 9138 and UGC 9556
would be $0^m.4$ lower than those used in the present paper.

The combined and calibrated images were utilized to obtain the radial
scale length $h$, vertical scale height
$z_0$, and "face-on" central surface brightness of the stellar disk, as
well as bulge-to-disk luminosity ratio $L_b/L_d$.

The images were rotated to align the galactic plane parallel to 
the horizontal axis. Choosing the rotation angle, we point our attention 
at the intermediate regions of galactic disks where a possible bulge does 
not reveal itself and the signal-to-noise ratio (S/N hereafter) is high 
enough.

We applied a standard method \citep{vdK81} to derive the structural 
parameters extracting photometric profiles parallel to the major and minor 
galactic axes. The radial scale length was obtained from two
photometric profiles extracted parallel to the major axis and displaced with
respect to the galactic midplane. 
This allows us to minimize the effects of dust absorption, because 
we avoid the galactic midplane. An average displacement is
of the order 0.7~$z_0$ (see below).
If the bulge was present, the central part
of radial photometric profiles (typically, 1 $h$ from the center)
is excluded from further analysis. 
We fit the function $f(r) ~=~
2 I_0 \, sech^2(z/z_0)\, \int^{R_{max}}_0 ~exp (-l / h) r ~dl$ to the 
radial profiles and
find mean values of $I_0$ and $h$. Here, $r$ is the distance to
the center, and $R_{max}$ is the distance to the edge of disk.
One can assume that $R_{max} = 4 h$ according to \citet{pohlen02,hm01}.
The integration was made along the line of sight $l$.
Two radial profiles utilized for the fitting are shown in the middle
panels of  Fig.~1--14 by the solid lines. Note that each a profile 
was manually cleaned from foreground stars before the fitting. 
The radial profile drawn through the galactic plane is shown in the
middle panels of Fig.~1--14 by the dashed line.

As a next step, we draw 10 to 14 cuts made parallel to the minor
axis of a galaxy and fit each photometric profile with the function
$f(z) ~=~ I~ sech^2 (|z + dz|/z_0)$. Here $|z|$ designates the distance
to the galactic plane. The "displacement term" $dz$ enables us to correct the
values of the disk scale height for a possible disorientation of major
axis or disk warp. The resulting value of the scale height $z_0$ and 
its error were found by averaging the values throughout the disk.
Our galaxies show no significant variations of the scale height with 
radius. Hence, we defined the mean scale height with no weights.

\begin{figure}
\epsscale{0.7}
\plotone{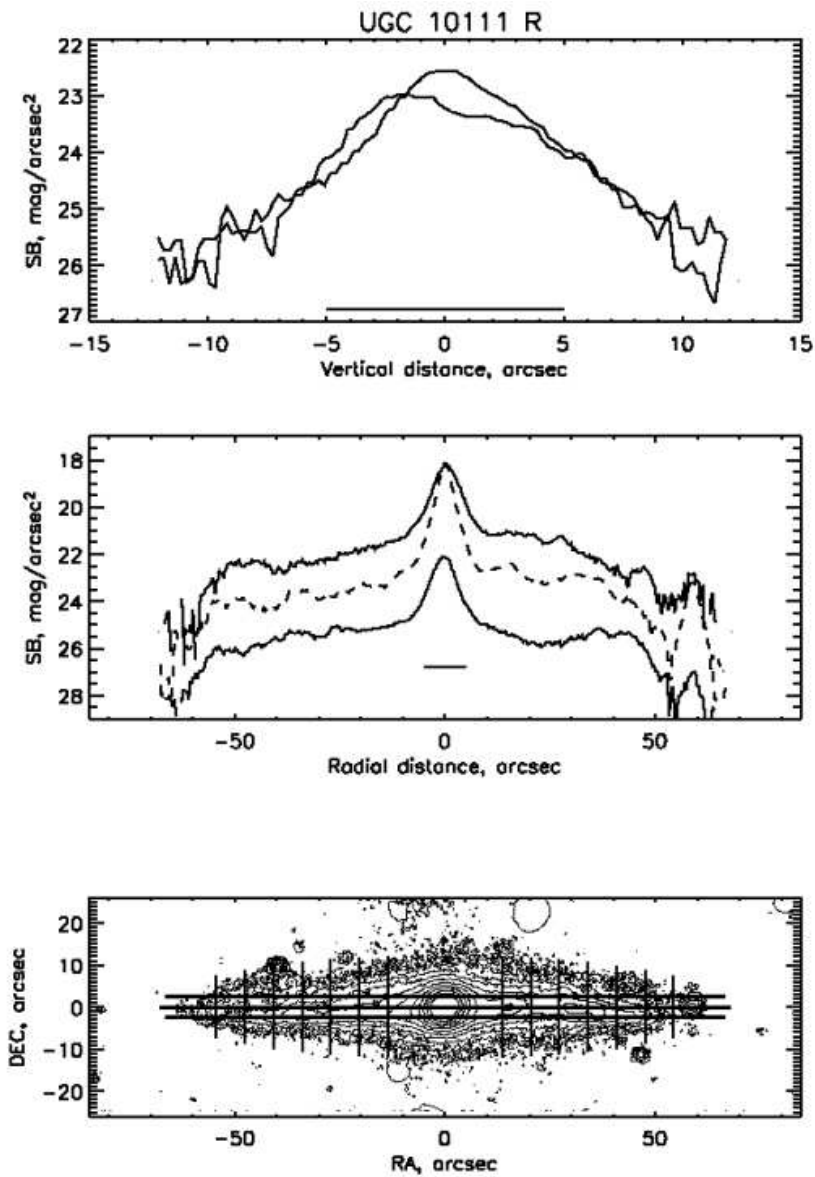}
\caption{\footnotesize
{\bf Upper:} Examples of the vertical profiles of UGC 10111 
extracted parallel to its minor axis. Both lines show profiles taken 
along two vertical cuts closest to the galactic center, see isophotal map
in the lower panel. The 10 arcseconds bar indicates the level of S/N=3.
{\bf Middle:} The radial profiles extracted along the major axis 
(dashed curve) and parallel to it (solid curves). The latter were used 
to derive the structural parameters of the galaxy. 
Upper and lower curves are shifted by +2 and -2 mag/arcsec$^2$ respectively 
from their real position. They are extracted along the upper and lower 
radial cuts shown on the isophotal map. The 10 arcseconds bar indicates 
the level of S/N=3.
{\bf Lower:} the isophotal map of UGC~10111. The isophotes correspond to 
20.5, 21, 21.5, 22, 22.5, 23, 23.5, 24, 24.5, 25, and 25.5 mag/arcsec$^2$ 
in the R band.
The places where the profiles were extracted are shown by lines. All 
artifacts and stars which can be seen in the picture were cleaned out 
manually before the structural parameters were found.
%
\label{fig1}}
\end{figure}

Fitting the profiles, we convolved the functions $f(z)$ with the gaussian 
smearing function assuming its FWHM from Table 1. The corresponding 
vertical profiles are shown in the upper frames of 
Fig.\ref{fig1}--\ref{fig1k}. They were manually cleaned of the 
foreground stars and artifacts before the fitting.

The value of the disk central surface brightness $\mu_0$
corrected to the face-on inclination was calculated with the parameters
$I_0$ and $z_0$ inferred above taking into account the photometric
calibration equations.
The extinction of our Galaxy (according to the LEDA database) is
also included into the analysis and listed in Table 2.


In order to check how examining only a limited number 
of brightness profiles (two radial and 10-14 vertical)
affects the inferred values, we derive the same values for each a galaxy
by extracting the radial profiles (drawn along the major axis)
with the increment of one pixel from 0.2 $z_{max}$ to 0.8 $z_{max}$ in the
vertical direction, where $z_{max}$ is the minor axis of an ellipse
encompassing the galaxy by the level of S/N=3. The vertical profiles in this
analysis were drawn with one-pixel increment taking a step off the disk
edge and its center. The resulting structural parameters are similar to
those obtained above using only a few photometric profiles. 
All conclusions of the paper remain unchanged in this case. 

As was shown by \citet{deG97}, we can neglect inclination corrections 
for inclinations larger than 86-87 degrees. Our V images
are deep enough to see obscuration by dust in most of our galaxies. 
Although dust is not seen in the galaxy FGC 1273, its bulge has a high 
degree of symmetry. Because its edge-on disk is
very thin, we assume that its inclination angle is 90\degr. 
For all other galaxies we can estimate the inclination angle of the 
galactic plane from the shape and positions of their dust layers and 
asymmetric position of center of brightness respective to external 
isophotes. The value of the inclination is shown in Table 3, its 
typical error is 0.7\degr. Based on those values, we applied no additional 
correction for non edge-on inclination to the photometric parameters 
derived above.

Assuming the inferred disk parameters, we subtracted the disk and extracted
the bulge images from the central parts of our galaxies. Then, the central 
parts of two radial profiles mentioned above, as well as the vertical 
profiles extracted along the minor axis, were utilized to estimate the 
bulge parameters. The King's profile $\rho_L^0 (1 + (r/a_b^k)^2)^{-3/2}$, 
as well as the exponential one $\rho_L^0 exp(-r/a_b^e)$, were used to fit 
the bulge volume luminosity density distribution.
Here, $\rho_L^0$ denotes the central volume luminosity density.
The bulge scales $a_b^k$ and $a_b^e$ could be different in the vertical and
radial directions (i.e. for oblate  bulges). The inner part of the vertical 
profiles were excluded from the analysis. 

Bulges of most galaxies in our sample are best fitted by the King's profile.
The only exception is UGC~9556, the bulge of which is best fitted by the 
exponential profile. Because the central part of the latter galaxy is oblate,
we suggest that that it probably has two disks: an HSB disk is encompassed 
by more extended LSB one. We consider its LSB disk throughout the paper. 
According to RFGC, UGC~9556 may have a lens in its central part. On the 
other hand, its type was defined as a galaxy with a bar (SB?c) in the UGC 
catalog \citep{UGC}. More over, UGC~9556 has an asymmetry of bright 
isophotes close to the galactic plane, which points toward a possible bar 
shielded by dust whose nerby side is seen. Indeed, photometric 
identification of bars in edge-on galaxies can rarely be conclusive.

With the help of the obtained parameters we infer the bulge-to-disk
luminosity ratio $L_b/L_d$. The main results of the fitting are 
shown in Table 3. The values of $h$ and $z_0$ are converted to the 
spatial units according to the adopted distances to the galaxies $D$. 
Table~3 also shows R-magnitudes and colors (V-R) derived for our objects. 
The magnitudes were obtained by integrating background-subtracted images 
of the galaxies within elliptical diaphragms. Major and minor axes of the 
diaphragms correspond to the sizes from the RFGC cataloge, which are 
quite similar to the galaxies' dimensions at S/N=3 level.


The distribution of $\mu_0$  (see Table 3) indicates the presence
of two subsamples: that with $\mu_0$  greater than 23.5
$mag/arcsec^2$, which we define as LSB galaxies, and that with a higher
surface brightness, which is designated as HSB galaxies in this paper. 
Hence, our sample consists of four LSB and seven HSB galaxies.
Note that all galaxies in the faintest RFGC surface brightness class 
were classified here as LSB objects.

Although our sample of objects enables to compare the structural
parameters of LSB and HSB disks, the sample is very limited. We incorporated
one more sample of edge-on galaxies whose photometric parameters have been
published by \citet{B94}. They made use of similar red photometric band and
technique to extract the photometric parameters.  We will utilize their data
together with ours throughout the paper in order to increase the
available sample of HSB galaxies. As it will be seen, the sample of 
Barteldrees \& Dettmar includes also one object, which can be classified 
as a LSB galaxy.

\section{LSB versus HSB: the vertical scale height of galactic disk
as a new feature to compare}

As was shown in \citet{B00,BM02,Resh03}, the galaxies of lower
surface brightness tend to have smaller $z_0/h$ ratios. However, this
conclusion was based on studies of mostly HSB galaxies. Now, we can
incorporate our LSB subsample and consider the relation between $z_0/h$
and the central surface brightness $\mu_0$. The objects from our sample
are denoted by squares in Fig. \ref{fig2}. The open squares show the HSB
subsample, whereas the filled ones designate LSB galaxies. The
galaxies taken from \citet{B94} are shown in Fig. \ref{fig2} with 
the crosses.

Furthermore, the near-infrared $K_s$-band sample of edge-on galaxies from
\citet{BM02} is available for comparison (the 2MASS sample hereafter). Here
we have to take into account the systematic difference in the brightness and
$z_0/h$ between the R and K photometric bands. As was noticed by
\citet{Z02}, the ratio of scales $z_0/h$ is 1.4 times less for the stellar
disk in K against R. It can be explained by stronger dust extinction 
in the R band, and was well illustrated by \citet{n891}. 
We corrected $z_0/h$ for the
2MASS galaxies taken from \citet{BM02} according to this value. The typical
color (R-K)=$2^m.1$ inferred for late-type face-on spirals by \citet{deJ96} 
was added to the central surface brightnesses of the 2MASS galaxies as well. 
The final correction that we applied was addition of the internal extinction
to the 2MASS central surface brightness, because it is low in the infrared
band and non-negligible in the R band. The value of this correction, 
$1^m.2$, is chosen so that the 2MASS sample coincides with our HSB objects
in Fig.\ref{fig2}.

Fig.\ref{fig2} shows all three samples together, where the 2MASS objects 
are denoted by the small filled triangles. A trend in Fig.\ref{fig2} is 
seen, an average difference of $2^m$ in $\mu_0$ leads to 1.5 change in 
the ratio of scales.
At the same time, there is no clear correlation found when $h$ and $z_0$
were plotted against $\mu_0$ separately. The correlation of $\mu_0$
versus $h$ was shown by \citet{Graham01}, but that conclusion was based on
mostly early-type spiral galaxies.

We also incorporate general galactic properties taken from the LEDA 
database into the analysis: absolute B magnitude $B_{abs}$, maximum of the 
rotation curve $V_m$, and HI index. The latter index denotes the difference 
between the B magnitude and the "HI magnitude". We found that LSB and HSB 
subsamples do not differ systematically in $B_{abs}$, $V_m$, and HI index. 
There is no correlation found between the values of $\mu_0$ and $z_0/h$ on 
the one hand, and $B_{abs}$, $V_m$, or HI index on the other hand.

In Fig.\ref{fig3} one can see a relation of Tully-Fisher type, 
where the values of the radial scale length are well correlated with the 
maximum of rotational velocity $V_m$.  According to 
\citet{Zwaan95,Sprayberry95,Chung02}, LSB and HSB spiral galaxies follow 
the same Tully-Fisher relation, and our Fig.\ref{fig3} is in a good 
agreement with this. 
It argues that we did not made a mistake deriving the spatial values.
Thus, the galaxy UGC 7808 was investigated by \citet{deG96} where the
shorter scale height value (in kpc) was inferred because of the lower adopted
distance to the galaxy. Fig.\ref{fig3} shows that our value of the scale 
length for the galaxy, 13.55 kpc, places the galaxy very close to the general
dependence in Fig.\ref{fig3}, whereas the scale length of 1.9 - 2.7 kpc taken
from \citet{deG96} would place this object far off. At the same time,
the angular values of the scale length found in the present work and in
the latter cited one, are very similar.

Following \citet{Z02}, we calculated the ratio of the total mass $M_t$ 
inside the optical radius to the luminosity of the galactic disk in the 
B band, $L_B$. We suppose that $M_t = G^{-1} 4h V_m^2$, 
where $G$ is the gravitational constant and $4 h$
radius encompasses the whole galaxy. The value of $L_B$ is obtained
from the absolute B-magnitude, which was taken from the LEDA and 
corrected for the internal galactic absorption. The values of
$M_t/L_B$ are plotted against the ratio $z_0/h$ in Fig.\ref{fig4}.

The notation in Fig.\ref{fig4} is the same as in Fig.\ref{fig2}. As was 
noticed by \citet{Z91}, the ratio of scales $z_0/h$ indicates the total 
mass of the spherical component of a galaxy expressed in its disk mass 
$M_s/M_d$. The relation between $z_0/h$ and $M_s/M_d$ obtained from 
numerical modeling (N body simulations) was published by \citet{Khop01} and 
shown in Fig.\ref{fig5}.  We made use of that dependence to evaluate
the model values of $M_s/M_d$ for our galaxies. 

Here, we have to clarify that we distinguish between a spherical and disk
subsystem throughout the paper. By the "spherical subsystem" we refer to both
a stellar bulge and dark halo, even if their shapes are not spherical but 
rather oblate (see discussion in section 5.5). In a general case, the spherical 
subsystem means a non-disk component, either stellar or not. The disk in 
our understanding is the galactic stellar disk. It consists mostly of stars 
for our objects. Later in the paper we also evaluate the ratio of 
dark-to-luminous masses. The dark mass belongs to the dark halo, 
whereas the luminous matter means the stellar bulge and disk.

Then, we take into account that $M_t = M_s + M_d$ and $L_B=
M_d / (M/L)$, where (M/L) denotes the B-band stellar mass-to-light ratio
in the disk. Hence $M_t/L_B = (M_s/M_d + 1)\cdot (M/L)$.
It is seen that the model value of $M_t/L_B$ depends on the adopted
B-band stellar mass-to-light ratio. 
The three curves in Fig.\ref{fig4} present the model 
values of $M_t/L_B$ which were calculated based on Fig.\ref{fig5} with 
the mass to light ratio (M/L) of 1, 5, and 15. As is seen in 
Fig.\ref{fig4}, most of the galaxies have values of (M/L) between 3 
and 10. The B-band stellar mass-to-light ratio (M/L) in 
Fig.\ref{fig4} corresponds to the distance taken along the horizontal axis 
toward the curve of (M/L)=1. The value of the stellar mass-to-light ratio 
is systematically higher for our LSB galaxies as compared to that of
HSB galaxies. 

This conclusion contradicts the bluer color of LSB galaxies found 
by \citep{deBlok95}
who give lower values for their (M/L), but the bulge-dominated LSB 
galaxies have colors comparable with HSB ones \citep{beijersbergen99}. 
The dereddened colors from our both LSB and HSB subsamples are almost 
the same (Table \ref{tab3}). On the other hand, LSB spirals have low 
metallicity as a rule. It might give the comparable colors,
whereas stellar disk's (M/L) takes larger values in LSB spirals. Another
reasonable explanation might be an excess of the dark matter in the 
disks of bulge dominated LSB spirals.

Large LSB galaxies have, on average, two times more mass in their gaseous 
component \citep{romanishin82} in comparison with HSB. 
Our LSB subsample has almost twice larger value of "HI index" against 
the HSB one. But this difference is not enough to explain the systematic 
difference in (M/L) in Fig.\ref{fig4} since the gas component does not 
dominate by mass in our galaxies.

The mass of the dark halo $M_h$ can be estimated from the relation
shown in Fig.4. The dark-to-luminous ratio 
is $(M_d + M_b)/M_h = (1+M_b/M_d) \cdot (M_d/M_h)$, 
where $M_b$ and $M_d$ denote masses of bulge and dark halo, respectively. 
On the other hand, $M_s/M_d = (M_h+M_b)/M_d$ and hence,
$M_h/M_d=M_s/M_d - M_b/M_d$. 
Combining previous equations, one can obtain:
\begin{equation}
\label{Mdark}
\frac{M_h}{M_d + M_b} = \frac{M_s/M_d - M_b/M_d}{1 + M_b/M_d} 
\end{equation}
The values of $M_b/M_d$ can be estimated from observations making a
rough assumption that the bulge-to-disk luminosity ratio follows the 
bulge-to-disk mass ratio $M_b/M_d = L_b/L_d$ (we consider how our 
conclusions might change for real galaxies where (M/L) is different 
for bulges and disks in section 5).
At the same time, $M_s/M_d$ can be estimated from Fig.\ref{fig5}.
The ratio of dark-to-luminous mass $M_h/(M_d + M_b)$ for our galaxies
is shown in Fig.\ref{fig6}. Surprisingly, there is no systematic difference
between the values of dark-to-luminous mass ratio for the galaxies with
different central surface brightnesses, see Fig.\ref{fig4}. It is generally
assumed that the LSB galaxies are dark-matter dominated, but all those
conclusions were based on studies of bulgeless galaxies. Our sample, on the
contrary, comprises mostly of the galaxies possessing non-negligible bulges.

We can also compare masses of the spherical subsystem $M_s$
(i.e. the sum of the bulge and halo) in our galaxies. In Fig.\ref{fig7} we
present how the spherical to disk mass ratio $M_s/M_d$ depends on the
disk central surface brightness. We kept the same notation as in
Fig.\ref{fig2} and Fig.\ref{fig4}. Fig.\ref{fig7} indicates that 
the LSB galaxies do not have more
massive dark matter halos. Instead, they have more massive
spherical subsystems. This supports a result by \citet{Graham02} that
not all LSB galaxies are dark matter dominated objects. Nevertheless, our 
result does not contradict previously made conclusions since the 
dark matter halo and the spherical subsystem become identical for 
bulgeless galaxies.

Differentiation between the bulge and halo allows us to demonstrate that 
there are dark-matter halo dominated large LSB galaxies as 
well LSB galaxies, the halos of which are less massive than their disks.

\section{Discussion}

\subsection{The sample selection}

There is no systematic difference in        
the obtained values of $\mu_0$ among our sample galaxies of I -- III surface       
brightness class (it was noticed by \citet{B00} as well). On the other    
hand, most galaxies of IV SB class are apparently bona fide LSB galaxies.
They constitute a small part of all RFGC objects (3\%). As was noticed 
by \citet{mcgaugh95}, there is a significant fraction of LSB galaxies 
with a large bulge to disk ratio. Bulges of LSB and HSB
systems are indistinguishable \citep{beijersbergen99}, yet their disks are  
different. Hence, one have to distinguish between LSB galaxies with and 
without bulges and take these possible bulges into account while 
undertaking a study of properties of dark halos in LSB galaxies,

Our paper does not attempt to present a statistically completed study of   
LSB spiral galaxies with large bulges. Instead, we compare two samples of 
objects of opposite properties. To make statistically reliable conclusions
the sample has to be extended.

\subsection{Selection effects}

Fig.\ref{fig2} presents a correlation between $\mu_0$ and 
$z_0/h$. Indeed, the values of $\mu_0$ and 
$z_0/h$ have not been obtained independently from each
other, as it follows from the formulae in section 3. Let us consider
how the non-90\degr\, inclination of the disk plane affects $\mu_0$
and $z_0/h$. If the inclination angle is less than 90\degr, the scale
height $z_0$ calculated in section 3 becomes overestimated.  At the
same time, the scale length $h$ is much less affected by the value
of inclination angle. On the other hand, the value of $z_0$ was taken into
account when the central surface brightness was calculated.
While overestimating the ratio $z_0/h$, we underestimate the disk
central surface brightness (hence, its numerical value will be larger).
It means that a non-90\degr inclination of disks shifts data
points in Fig.\ref{fig2} toward the right upper corner. 
Hence, the systematic errors due to inclination may only scatter the 
dependence shown in Fig.\ref{fig2} (say, for the 2MASS galaxies) and 
do not explain a good correlation.

The second effect that has to be considered is the internal dust absorption 
in galaxies. According to \citet{Xilouris99}, the scale length in 
dusty disks appears higher because of the scattering and absorption 
effects. On the other hand, the dust absorption decreases the derived 
central surface brightness. In our case it would shift data points 
in Fig.\ref{fig2}
from the upper left to the lower right corner and would form a dependence 
similar to that seen in Fig.\ref{fig2}. 
Nevertheless, we decreased
the influence of dust by avoiding the dust layer when extracting the radial
profiles. This allowed us to minimize the dust absorption. 
Furthermore, one can see that
the infrared and optical subsamples follow the similar dependence in
Fig.\ref{fig2}. This argues that the internal absorption has little effect on
the difference between LSB and HSB photometric parameters
and Fig.\ref{fig2} has a physical meaning.

\subsection{Internal absorption in galactic disks and the ratios $M_s/M_d$
and $M_h/(M_b + M_d)$}

As was noted in section 4, the disk thickness is different when it
is estimated in different photometric bands. All the considered relations
between the mass of a dark halo, stellar disk, and spherical
component are made using the data taken in the R band. On the other hand,
the infrared ratios of photometric scales $z_0/h$ are less than the
optical ones. The infrared values are more preferable because of the lower 
dust absorption, so we could decrease all our ratios $z_0/h$ by a factor
of 1.4. As is seen in Figs.\ref{fig4} and \ref{fig5}, a
proportional decrease of the scale ratio affects Figs.\ref{fig6} 
and \ref{fig7} only quantitatively. Hence, all previous conclusions 
remain unchanged.

The values of $\mu_0$ inferred for our HSB galaxies are less than the
Freeman's value (taking into account a difference between the R and B 
bands). It implies that the internal extinction
may be important in the disks of our galaxies.
Since all the galaxies are spiral and are relatively nearby, one can assume
roughly the same dust-to-stars ratio in them. Then, the internal 
extinction proportionally increases the values of $\mu_0$.
At the same time, it does not change the main trends in Fig.\ref{fig2}, 
\ref{fig4}, \ref{fig6}, and \ref{fig7}.

In a more complicated case, the internal dust extinction 
may be systematically different in the galaxies of our sample. 
Thus, according to \citep{mcgaugh94,Matt01} LSB spirals are
likely to be less dusty than HSB ones. One can see that it strengthens 
the relation shown in Fig.\ref{fig2}:
extinction correction of $\mu_0$ for HSB spirals moves data points 
further to the left than it does for LSB spirals. As a result, we can always 
distinguish between these two subsamples. It corresponds to the 
conclusion made by \citet{beijersbergen99} that the dust extinction 
alone can not explain the difference in surface brightness between 
LSB and HSB spirals. 

Another way to take the extinction into account is to connect it with 
the global galactic parameters such as the absolute magnitude or 
rotational velocity, see \citep{tully98} and references therein. 
Correction of $\mu_0$ for the extinction with the help of absolute 
R-magnitudes or $V_m$ moves data points to the left in Fig.\ref{fig2} 
and does not change its general trend.

The bulge-to-disk luminosity ratio has been utilized to draw 
Fig.\ref{fig2} and \ref{fig6}. Since part of our galaxies has
bulges, attention should be paid to how the extinction may change the
derived values of $L_b/L_d$. In addition to the profile fitting, we
conducted a direct integration of bulges. At first, the model disk
(constructed according to the parameters defined during the disk fitting) 
was subtracted from the images of galaxies. 
Then, we integrated all the central
part which was above the zero level. The ratio of the integrated 
luminosity of the bulge to the model disk luminosity  
$L_b^I/L_d$ gives us a lower
bound of $L_b/L_d$ ratio (because the model disk is "dust-free", 
and the bulge is dimmed by the extinction). The value of $L_b^I/L_d$ is 
2-6 times lower than the value of  $L_b/L_d$ given in Table \ref{tab3}. 
If we use $L_b^I/L_d$ instead of $L_b/L_d$, Fig.~\ref{fig2} does not
change qualitatively. 
On the other hand, in Fig.\ref{fig6} all our LSB galaxies move to the right, 
since a larger fraction of mass of their spherical component is assigned 
to the dark halo. Then, if we apply $L_b^I/L_d$ as a bulge-to-disk
luminosity ratio, one cannot conclude that the 
ratio "(dark halo + bulge)/disk" in the bulge-dominated galaxies is 
systematically higher whereas dark-to-luminous ratio not. 
In this case dark-to-luminous mass ratio would be higher in our LSB 
systems too.

Alternatively, one can obtain the ratio $L_b/L_d$ from direct
integration of the bulge and disk from our images. In contrast to the 
previous case that gives the lower limit on $L_b/L_d$, this integration 
yields values of $L_b/L_d$ that are systematically higher than it can be 
seen in Table \ref{tab3}. This method of evaluation of bulge-to-disk
luminosity ratio does not change our conclusions as well.

\subsection{M/L may be different for bulge and disk}

Assuming that $M_b/M_d = L_b/L_d$, one can notice that indeed,
bulges and disks have different colors and, hence, their 
stellar population has to show different mass-to-light ratios. It 
does not affect all our results except for Eq.~(\ref{Mdark}) and 
Fig.~\ref{fig6}. Bulges are redder than disks as a rule 
\citep{peletier96} and have larger M/L. Then, $M_h/(M_d + M_b)$ is 
overestimated for galaxies with significant bulges (LSB galaxies 
in our sample). Hence, it supports our conclusion that dark halo 
does not dominate in LSB galaxies which have big bulges. 
It should be noticed 
that the difference between colors of bulges and disks is very 
small \citep{peletier96,gadotti01}, which makes the effect mentioned 
above insignificant.


\subsection{Oblate bulges, non-spherical halos}

Dark matter halos and bulges of galaxies may not be exactly
spherical, but rather oblate. Our definition of $M_t = G^{-1}4hV_m^2$
works well for the case of spherical symmetry. In a general case 
$M_t = \eta G^{-1}4hV_m^2$, where $\eta$ is a dimensionless parameter,
the value of which is determined by the mass distribution, and
$\eta < 1$ for the case of galaxies. If the whole mass of a galaxy 
was enclosed in a thin exponential disk, the parameter $\eta$ taken 
at $4h$ distance from the center is approximately equal 
to 0.5 \citep{freeman70}. All other reasonable geometric cases represent 
a mixture of disk and spherical components and give $\eta$ between 0.5 
and 1. For the case of non-spherical dark halo, the difference 
between LSB and HSB galaxies in Fig.\ref{fig4} (and, hence, in M/L for 
the stellar disk) would be even more prominent, because $M_t$ calculated 
using $\eta$ is systematically lower for disk-dominated HSB spirals
than for bulge-dominated LSB galaxies in our sample. Note that once a 
non-disk component is presented in all our galaxies, the difference 
in $\eta$ be significantly less than a factor of 2. 
At the same time, it does not change other conclusions of this paper.

A possible existence of a non-spherical, oblate component was not taken 
into account by \citet{Khop01}. If one takes it into account, the general 
trend shown in Fig.~\ref{fig4} remains unchanged. However, a systematic 
difference between the ellipticity of dark halos in LSB and HSB galaxies 
can significantly affect Fig.~\ref{fig4}. For instance, an assumption of 
oblate halo in LSB spirals and spherical halo in HSB ones decreases the 
difference between the B-band mass-to-luminosity ratio in stellar disks 
mentioned above, since it shifts data points to the left (though, by less 
than a factor of 2). On the other hand, we show that the dark halos are 
likely to be not too massive and, hence, not dominant by mass in our 
bulge-dominated LSB spirals. Therefore, the role of their non-spherical 
shapes is insignificant. It is doubtful that the systematic difference 
between the ellipticity of dark halos in LSB and HSB galaxies can affect 
our conclusions. 
Furthermore, if a significant fraction of dark matter in the
bulge-dominated LSB galaxies is located in their disks, it helps to rise 
their (M/L) as it can be seen from Fig.\ref{fig4}. 
Note that one of candidates to the dark matter, namely cold 
molecular clouds, could form a disk-like subsystem \citep{pfenniger94}.

\section{Conclusion}

1) We present results of photometric observations of a sample of
edge-on galaxies. Our sample includes four LSB and seven HSB galaxies.
The photometric disk scales (both vertical and radial), 
disk central surface brightness, and bulge-to-disk luminosity ratios 
were derived.

2) Stellar disks of LSB galaxies are thinner (when parameterized by the
ratio $z_0/h$) than HSB ones. There is a
clear correlation between their central surface brightnesses and the
vertical to radial scale ratios. 

3) While having different central surface brightnesses and bulge-to-disk 
ratios, the LSB and HSB galaxies in our sample follow the same 
dependence "disk scale length versus the maximum rotational 
velocity".

4) Our LSB galaxies tend to harbor massive spherical subsystems
(bulge + halo) as well as to have higher values of the
mass-to-luminosity ratio in their disks when compared to the HSB
objects. Nevertheless, the dark halo is not strictly the most massive
subsystem in our bulge-dominated LSB galaxies. 
The LSB spirals appear to be the "spherical subsystem dominated" galaxies
but not always the "dark matter dominated".

\begin{acknowledgments}
D.B. is supported by NASA/JPL
through the grant 99-04-OSS-058. The project was partially supported by
Russian Foundation for Basic research via the grant 04-02-16518.
We have made use of the LEDA database. We thank Verne Smith 
and Michael Endl for their comments on the manuscript and the 
anonymous referee whose commentaries and corrections essentially 
improved the paper. D.B. is grateful to A.Khoperskov and Eduard Vorobyov 
for fruitful discussions and help.

\end{acknowledgments}

\clearpage

\begin{table}[h]
\begin{center}
\caption{Summary of the observing run \label{tab1}}
\begin{tabular}{lccrrcccc}
\tableline\tableline
 Name & SB class & Band &  Date & Int. time & Nexp & Seeing & Sky & S/N=3 level \\
      &          &      &  1999 & sec.      &      & arcsec & mag/arcsec$^2$ & $mag/arcsec^2$ \\
\tableline\tableline
UGC 10111 & IV  & V & 27 Apr    &  600  & 1 & 1.9 & 21.37 & 26.58 \\
          &     & R & 27 Apr    & 1200  & 4 & 2.0 & 20.72 & 26.77 \\
UGC 11301 & III & V & 27 Apr    &  700  & 3 & 1.6 & 20.76 & 25.17 \\
          &     & R & 27 Apr    & 1000  & 4 & 1.6 & 20.38 & 25.81 \\
UGC 5662  & III & V & 30 Apr    &  600  & 1 & 3.0 & 21.39 & 26.64 \\
          &     & R & 30 Apr    & 1200  & 2 & 2.4 & 20.59 & 26.63 \\
UGC 6080  &  II & V & 30 Apr    &  600  & 1 & 1.9 & 21.49 & 26.46 \\
          &     & R & 30 Apr    & 1200  & 2 & 1.7 & 20.68 & 26.60 \\
UGC 6686  & III & V & 27 Apr    & 1200  & 2 & 1.8 & 21.35 & 26.85 \\
          &     & R & 27 Apr    &  900  & 3 & 1.7 & 20.48 & 26.45 \\
UGC 7808  &  IV & V & 27 Apr    &  600  & 1 & 2.2 & 21.35 & 26.40 \\
          &     & R & 27 Apr    & 1200  & 2 & 2.0 & 20.56 & 26.66 \\
UGC 9138  &   I & V & 28 Apr    &  900  & 2 & 1.1 & 21.37 & 26.48 \\
          &     & R & 28 Apr    &  900  & 2 & 1.0 & 20.63 & 26.90 \\
UGC 9422  &   I & R & 30 Apr    & 1200  & 2 & 1.7 & 20.57 & 26.91 \\
UGC 9556  &  IV & V & 28 Apr    & 1800  & 4 & 1.0 & 21.46 & 26.51 \\
          &     & R & 28 Apr    & 2900  & 6 & 1.0 & 20.68 & 27.26 \\
FGC 1273  &  IV & V & 27 Apr    &  600  & 2 & 1.8 & 21.43 & 26.54 \\
          &     & R & 27 Apr    &  900  & 2 & 1.7 & 20.50 & 26.59 \\
NGC 4738  &   I & R & 30 Apr    & 1200  & 2 & 1.5 & 20.45 & 26.56 \\
\tableline
\end{tabular}
\end{center}
\vbox{
Name of galaxy, surface brightness class (according
to the RFGC catalog), photometric band, date of observations, 
exposure time (total), number of expositions, average seeing, 
level of sky brightness, and S/N=3 SB level in combined frames.}

\end{table}


\begin{table}[t]
\begin{center}
\caption{General galactic parameters utilized in the paper \label{tab2}}
\begin{tabular}{lrllllll}
\tableline\tableline
Name      & D,Mpc & Type &$logD_{25}$& $b_t$ & $A_B$ & $logV_m$ & $B_{abs}$ \\
\tableline\tableline
UGC 10111 & 139.6 & Sc   & 1.221 & 16.  & 0.178 & 2.370 & -21.3 \\
UGC 11301 &  62.3 & Sc   & 1.295 & 15.5 & 1.273 & 2.379 & -21.2 \\
UGC 5662  &  17.1 & SBb  & 1.495 & 15.4 & 0.115 & 1.899 & -17.6 \\
UGC 6080  &  30.3 & Scd  & 1.3   & 15.8 & 0.036 & 1.877 & -18.6 \\
UGC 6686  &  86.4 & Sb   & 1.418 & 15.0 & 0.135 & 2.283 & -21.2 \\
UGC 7808  &  96.3 & Sb   & 1.492 & 14.6 & 0.098 & 2.403 & -21.8 \\
UGC 9138  &  61.9 & Sc   & 1.284 & 14.8 & 0.108 & 2.161 & -20.9 \\
UGC 9422  &  45.6 & Sc   & 1.279 & 14.7 & 0.1   & 2.140 & -20.5 \\
UGC 9556  &  32.5 & SBc  & 1.099 & 16.0 & 0.043 & 1.974 & -18.1 \\
FGC 1273  &  49.4 & Sc   & 0.801 & 16.5 & 0.103 & 2.166 & -18.0 \\
NGC 4738  &  63.6 & Sc   & 1.297 & 14.3 & 0.076 & 2.335 & -21.4 \\
\tableline
\end{tabular}
\smallskip
\end{center}
\vbox{
Name of galaxy, adopted distance
(corresponding to the Hubble constant H$_0$ = 75 km~s$^{-1}$~Mpc$^{-1}$),
morphological type, the major axis size $logD_{25}$ (in 0.1\arcmin), 
total B-band magnitude, foreground extinction in our Galaxy in B, 
logarithm of rotational velocity $log~V_m$, and absolute 
B-magnitude (all those values are taken from the LEDA).
}
\end{table}

\begin{table}
\begin{center}
\caption{The derived structural parameters of the galaxies \label{tab3}}
\begin{tabular}{lcrccrccrr}
\tableline\tableline
Name &i   & PA& $h$ & $z_0$& $z_0/h$ & $\mu_0$        & $L_b/L_d$ & $m_R$ & V-R\\
     &deg &deg&kpc & kpc  &         & mag/arcsec$^2$ &           &    mag & mag  \\
\tableline\tableline
UGC 10111 & 88.2 &37.5 & 15.84 $\pm$ 0.14 & 2.60 $\pm$ 0.42 & 0.168 & 24.63 $\pm$ 0.11 & 0.58 & 15.08 & 0.63 \\
UGC 11301 & 88.2 &110. &  8.24 $\pm$ 0.84 & 1.30 $\pm$ 0.08 & 0.160 & 22.12 $\pm$ 0.13 & 0.25 & 13.01 & 0.81 \\
UGC 5662  & 89.3 &147.5&  2.16 $\pm$ 0.51 & 0.50 $\pm$ 0.06 & 0.237 & 22.70 $\pm$ 0.37 & 0.21 & 13.62 & 0.61 \\
UGC 6080  & 86(?)&125. &  2.93 $\pm$ 0.18 & 0.70 $\pm$ 0.05 & 0.236 & 22.69 $\pm$ 0.16 & 0.00 & 14.45 & 0.50 \\
UGC 6686  & 88.0 &50.  &  9.82 $\pm$ 1.27 & 1.89 $\pm$ 0.21 & 0.196 & 22.51 $\pm$ 0.13 & 0.10 & 13.44 & 0.71 \\
UGC 7808  & 88.7 &93.5 & 13.55 $\pm$ 2.63 & 1.44 $\pm$ 0.18 & 0.158 & 23.97 $\pm$ 0.18 & 0.86 & 13.60 & 0.69 \\
UGC 9138  & 87.0 &168.5&  4.71 $\pm$ 0.12 & 1.05 $\pm$ 0.06 & 0.214 & 22.42 $\pm$ 0.38 & 0.07 & 14.79 & 0.76 \\
UGC 9422  & 88.3 &159. &  3.54 $\pm$ 0.13 & 0.80 $\pm$ 0.05 & 0.225 & 22.64 $\pm$ 0.35 & 0.00 & 14.96 & --- \\
UGC 9556  & 87.0 &133. &  3.63 $\pm$ 0.49 & 0.54 $\pm$ 0.09 & 0.141 & 24.77 $\pm$ 0.60 & 1.00 & 15.39 & 0.39 \\
FGC 1273  & 89.7 &170. &  5.17 $\pm$ 0.98 & 0.63 $\pm$ 0.10 & 0.118 & 24.33 $\pm$ 0.18 & 1.37 & 14.70 & 0.68 \\
NGC 4738  & 85.7 &32.5 &  4.68 $\pm$ 0.21 & 1.23 $\pm$ 0.08 & 0.270 & 21.11 $\pm$ 0.16 & 0.01 & 13.39 & ---\\
\tableline
\end{tabular}
\end{center}
\vbox{
Parameters derived for our galaxies: inclination angle, position angle,
disk scale length in kpc, disk scale height, ratio of scales $z_0/h$,
stellar disk central surface brightness, bulge to disk ratio, R-magnitude 
integrated within the elliptical diaphragm with axes size taken from
the RFGC catalog, and color (V-R). The R-magnitude and the color are 
uncorrected for the foreground extinction.
}
\end{table}


\clearpage


\clearpage
\begin{figure}
\epsscale{0.9}
\plotone{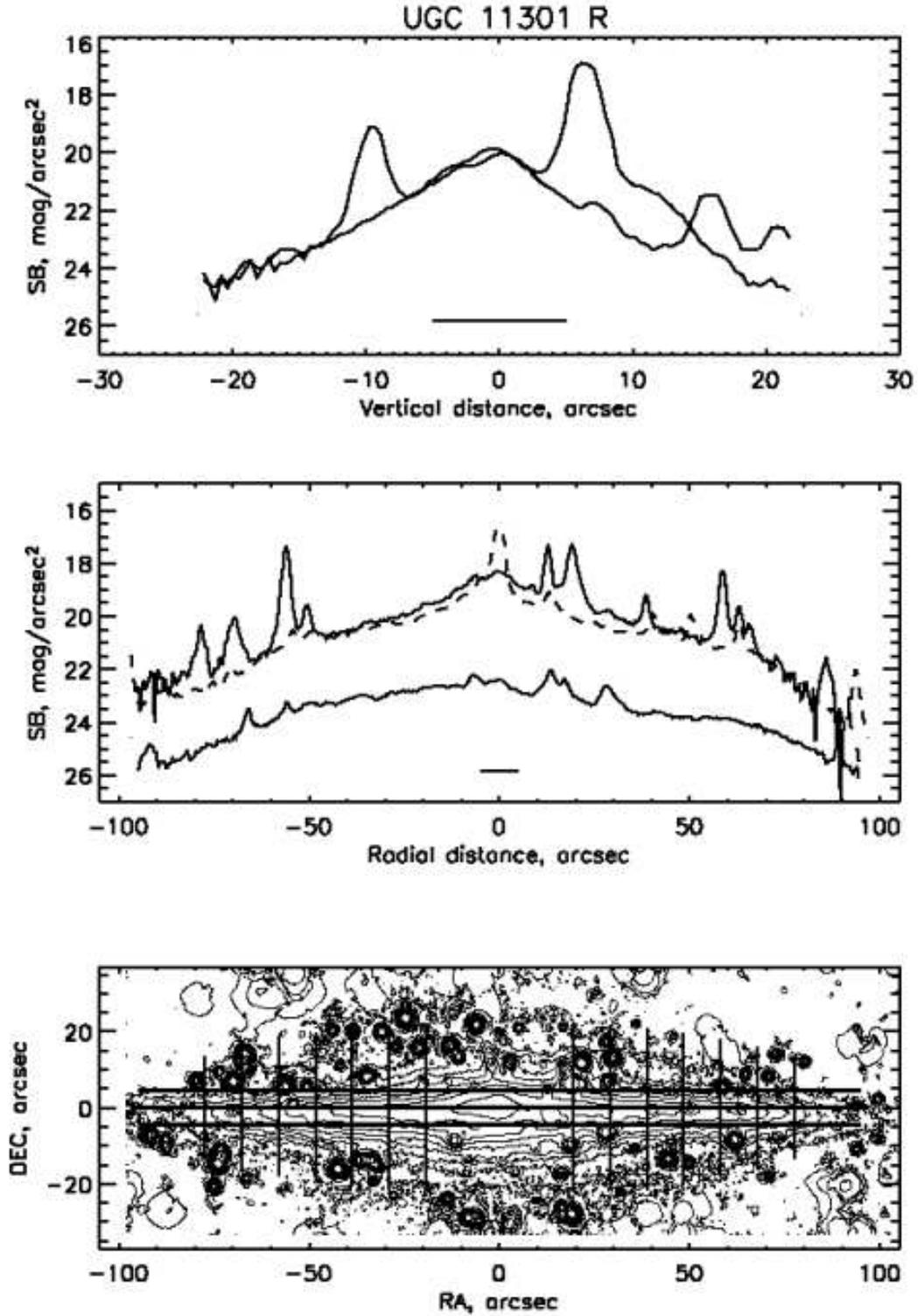}
\caption{
The same for UGC~11301. The isophotes correspond to
19.5, 20.5, 21, 21.5, 22, 22.5, 23 ,23.5, and 24.5 mag/arcsec$^2$.
}
\end{figure}


\clearpage
\begin{figure}
\epsscale{0.9}
\plotone{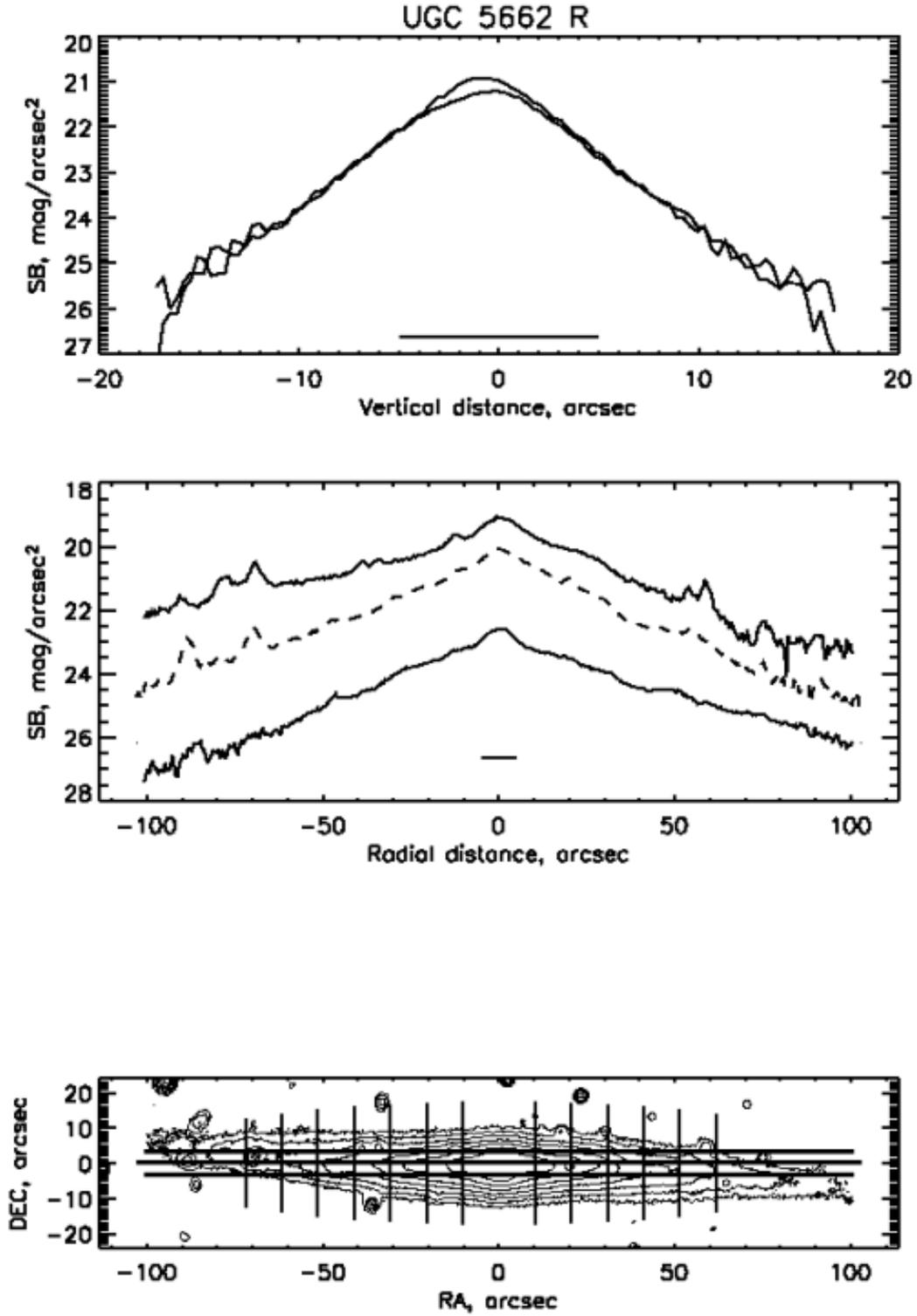}
\caption{
The same for UGC~5662. The isophotes correspond to 
20.5, 21, 21.5, 22, 22.5, 23, 23.5, and 24
mag/arcsec$^2$.
}
\end{figure} 


\clearpage
\begin{figure}
\epsscale{0.9}
\plotone{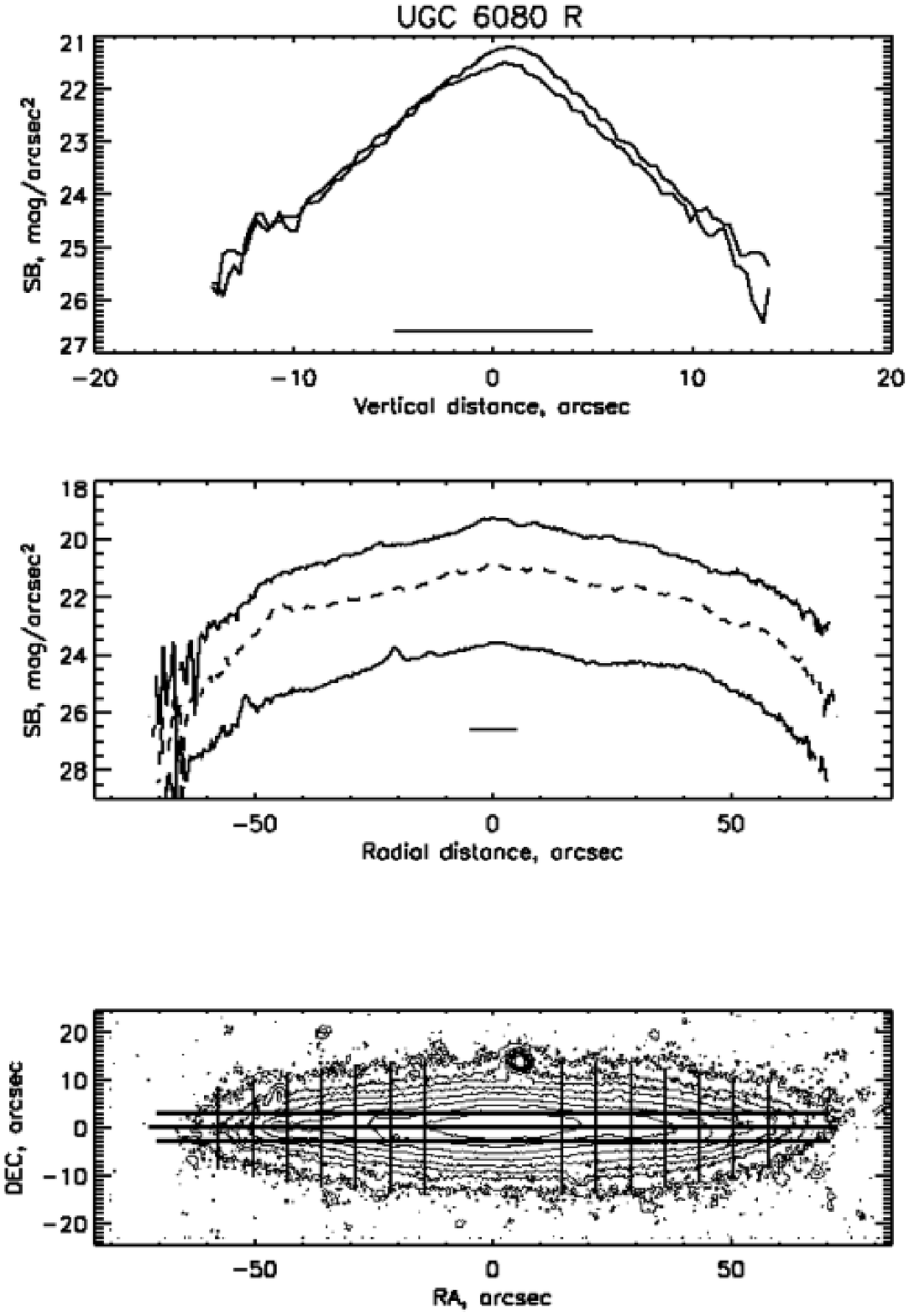}
\caption{
The same for UGC~6080. The isophotes correspond to    
20.5, 21, 21.5, 22, 22.5, 23, 23.5, 24, 24.5, and 25.5
mag/arcsec$^2$.
}

\end{figure}


\clearpage
\begin{figure}
\epsscale{0.9}
\plotone{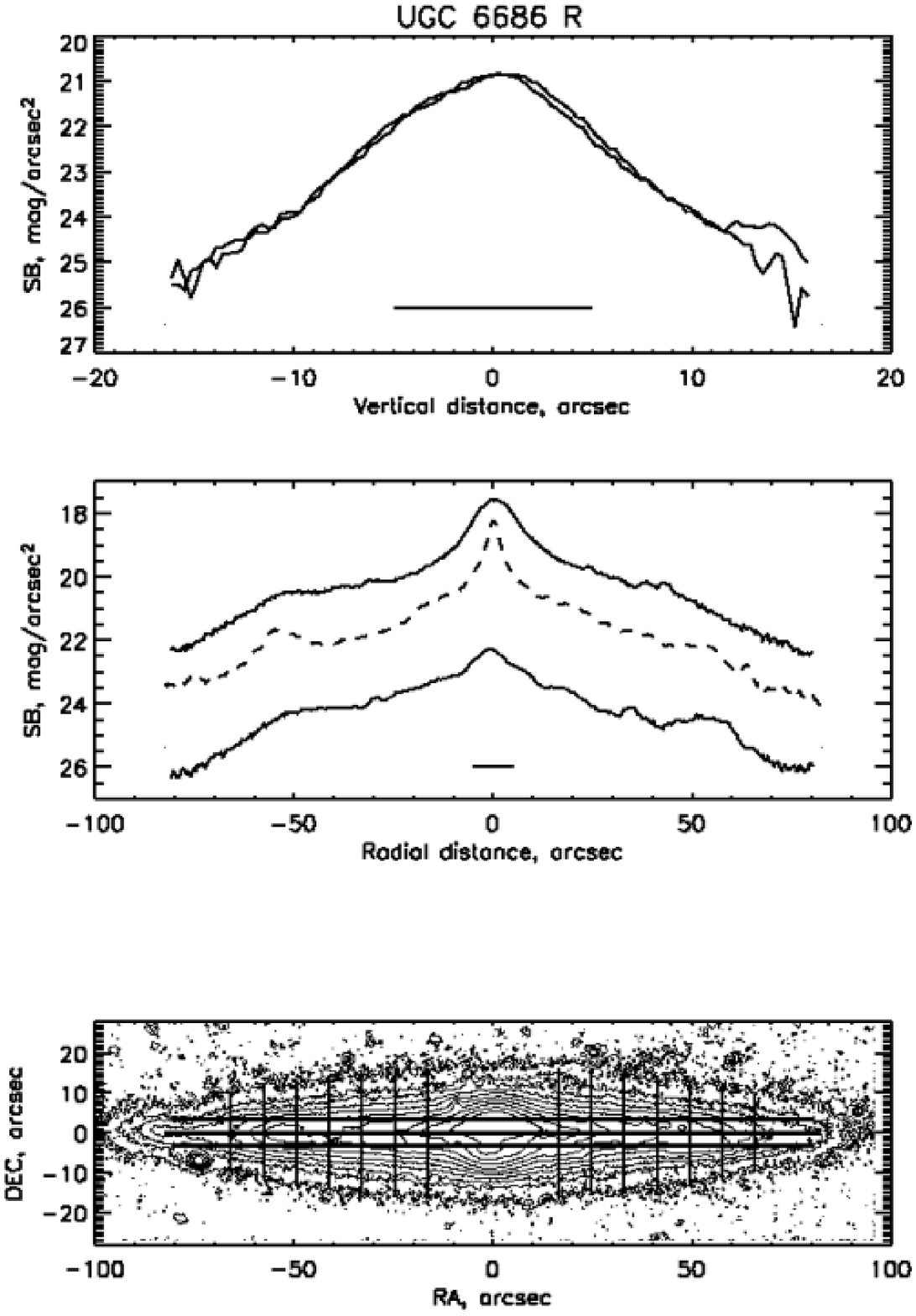}
\caption{
The same for UGC~6686. The isophotes correspond to    
20.5, 21, 21.5, 22, 22.5, 23, 23.5, 24, 24.5, and 25.5
mag/arcsec$^2$.
}

\end{figure} 


\clearpage
\begin{figure}
\epsscale{0.9}
\plotone{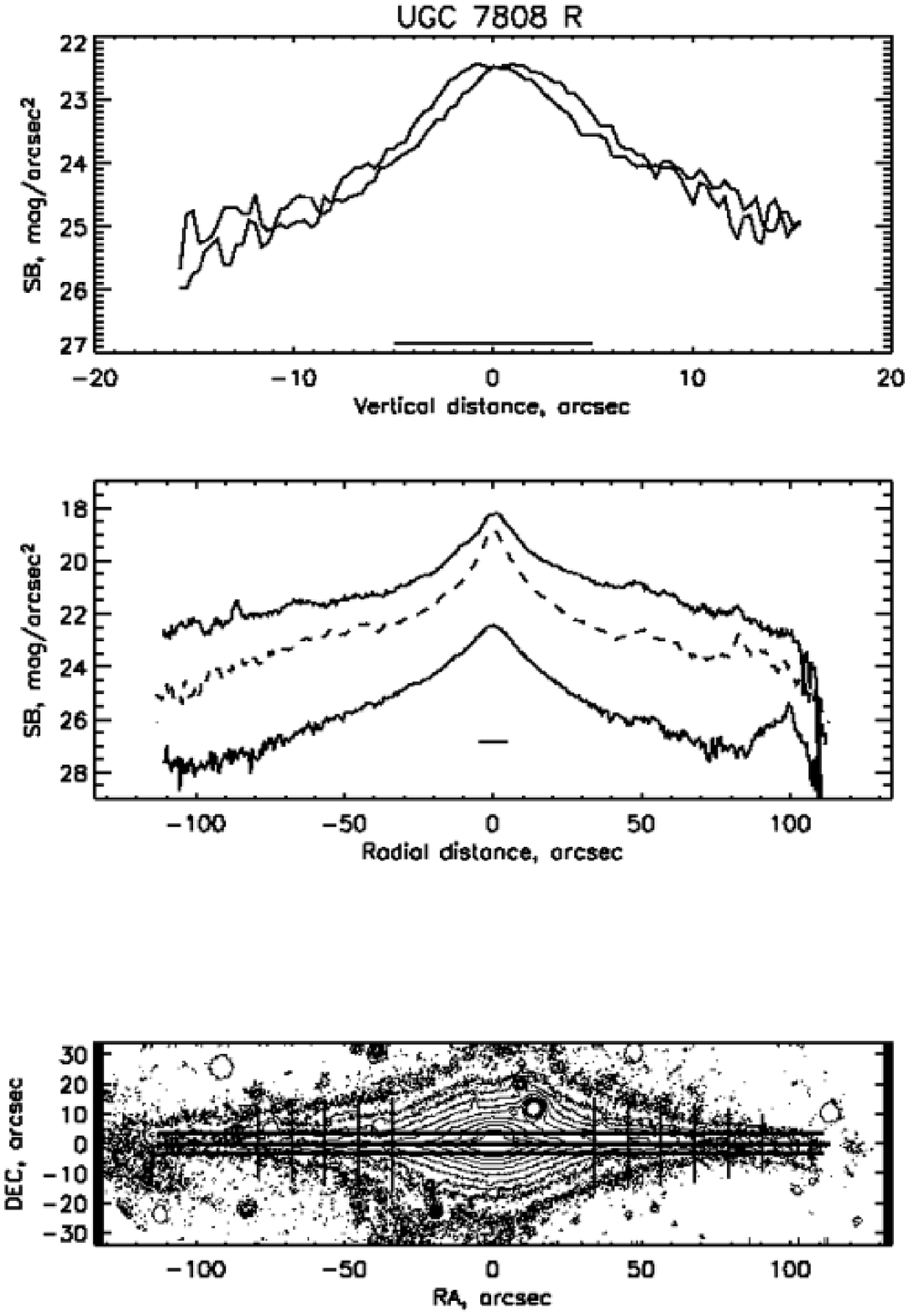}
\caption{
The same for UGC~7808. The isophotes correspond to    
20.5, 21, 21.5, 22, 22.5, 23, 23.5, 24, 24.5, and 25.5
mag/arcsec$^2$.
}

\end{figure}


\clearpage
\begin{figure}
\epsscale{0.9}
\plotone{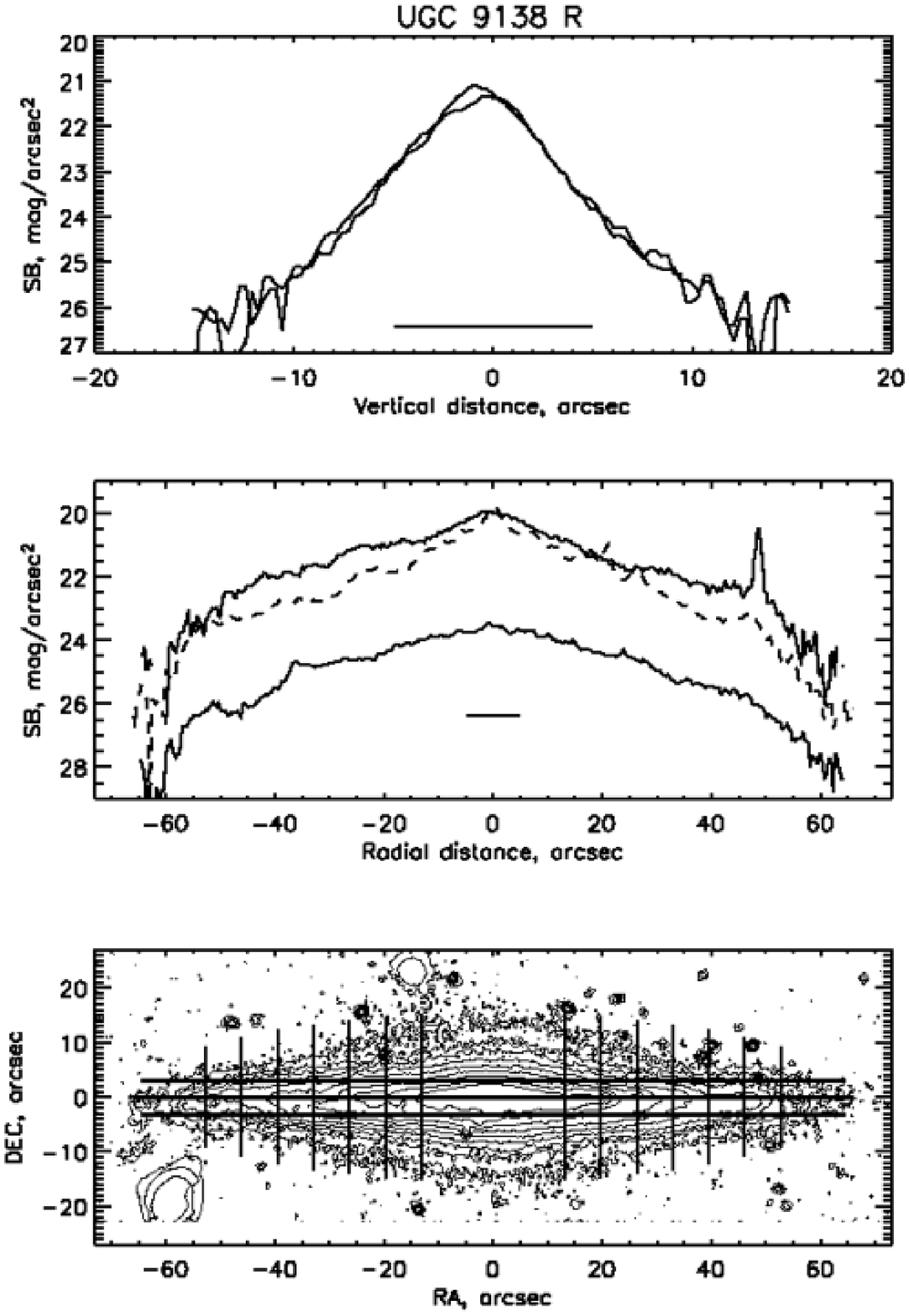}
\caption{
The same for UGC~9138. The isophotes correspond to    
20.5, 21, 21.5, 22, 22.5, 23, 23.5, 24, 24.5, 25, and 26
mag/arcsec$^2$.
}

\end{figure} 


\clearpage
\begin{figure}
\epsscale{0.9}
\plotone{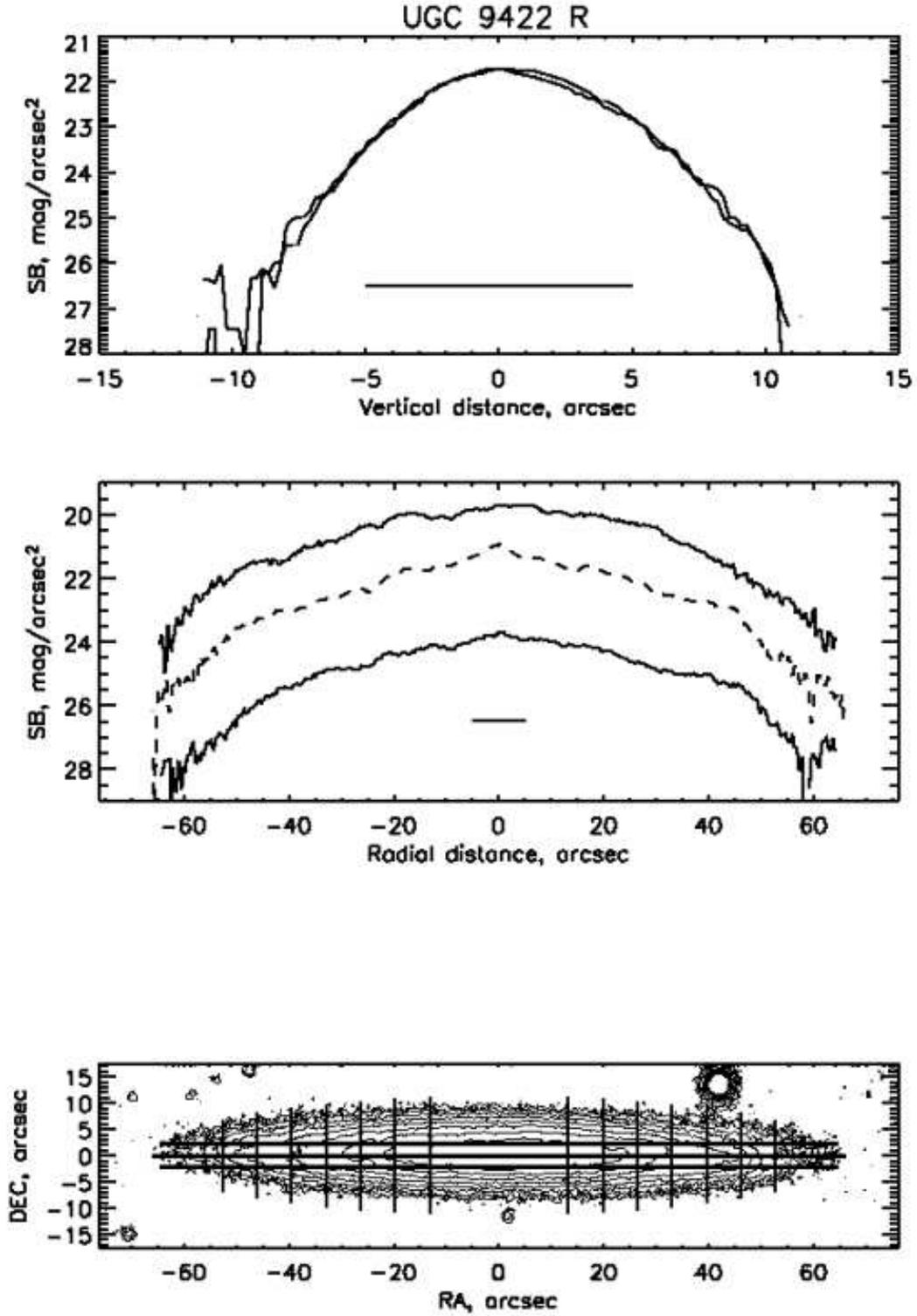}
\caption{
The same for UGC~9422. The isophotes correspond to    
20.5, 21, 21.5, 22, 22.5, 23, 23.5, 24, 24.5, 25,25.5
mag/arcsec$^2$.
}

\end{figure}


\clearpage
\begin{figure}
\epsscale{0.9}
\plotone{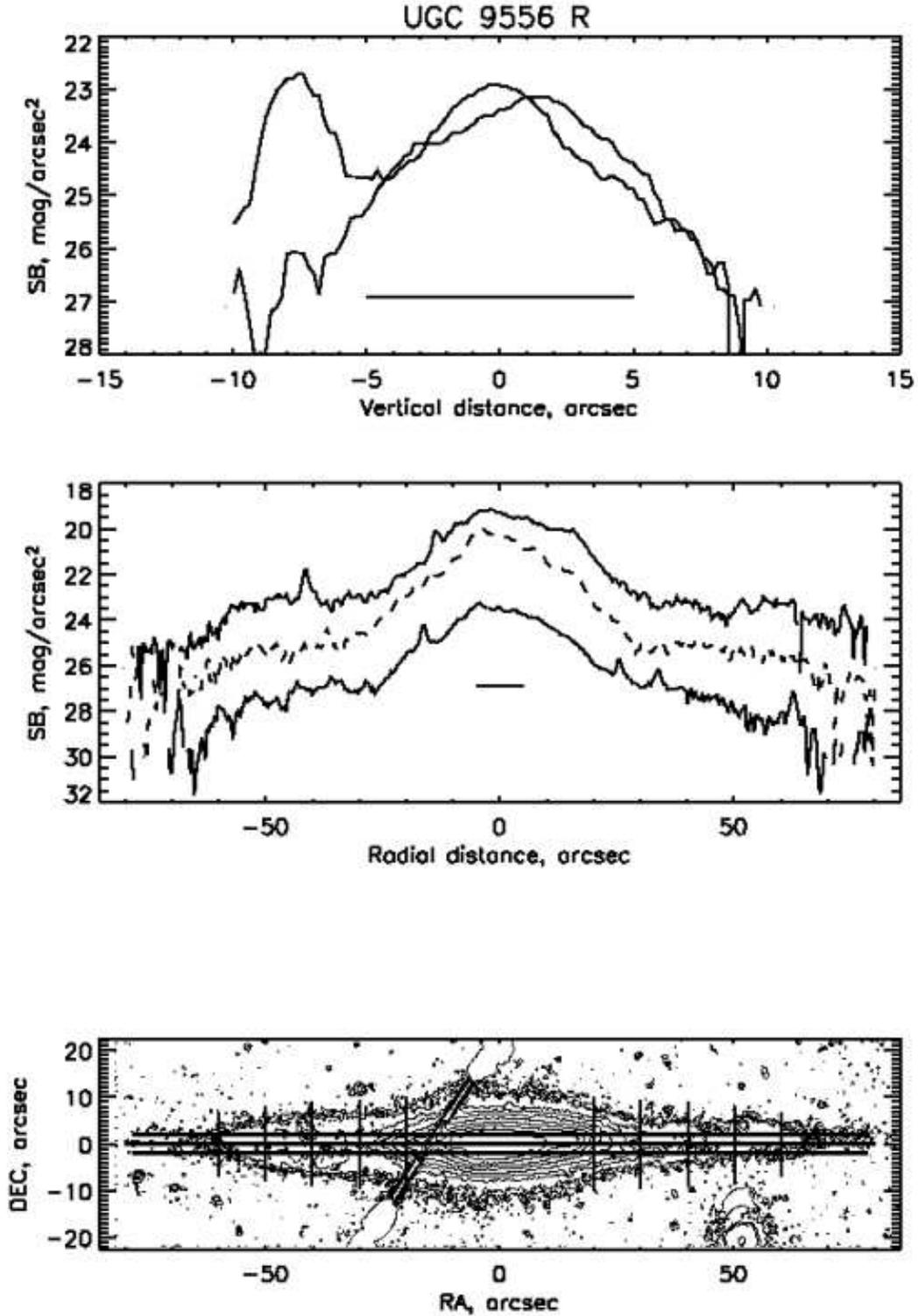}
\caption{
The same for UGC~9556. The isophotes correspond to    
20.5, 21, 21.5, 22, 22.5, 23, 23.5, 24, 24.5, 25, 26, and 26.7
mag/arcsec$^2$. The diagonal feature in the lower panel is a 
remainder of a bright satellite track.
}

\end{figure} 


\clearpage
\begin{figure}
\epsscale{0.9}
\plotone{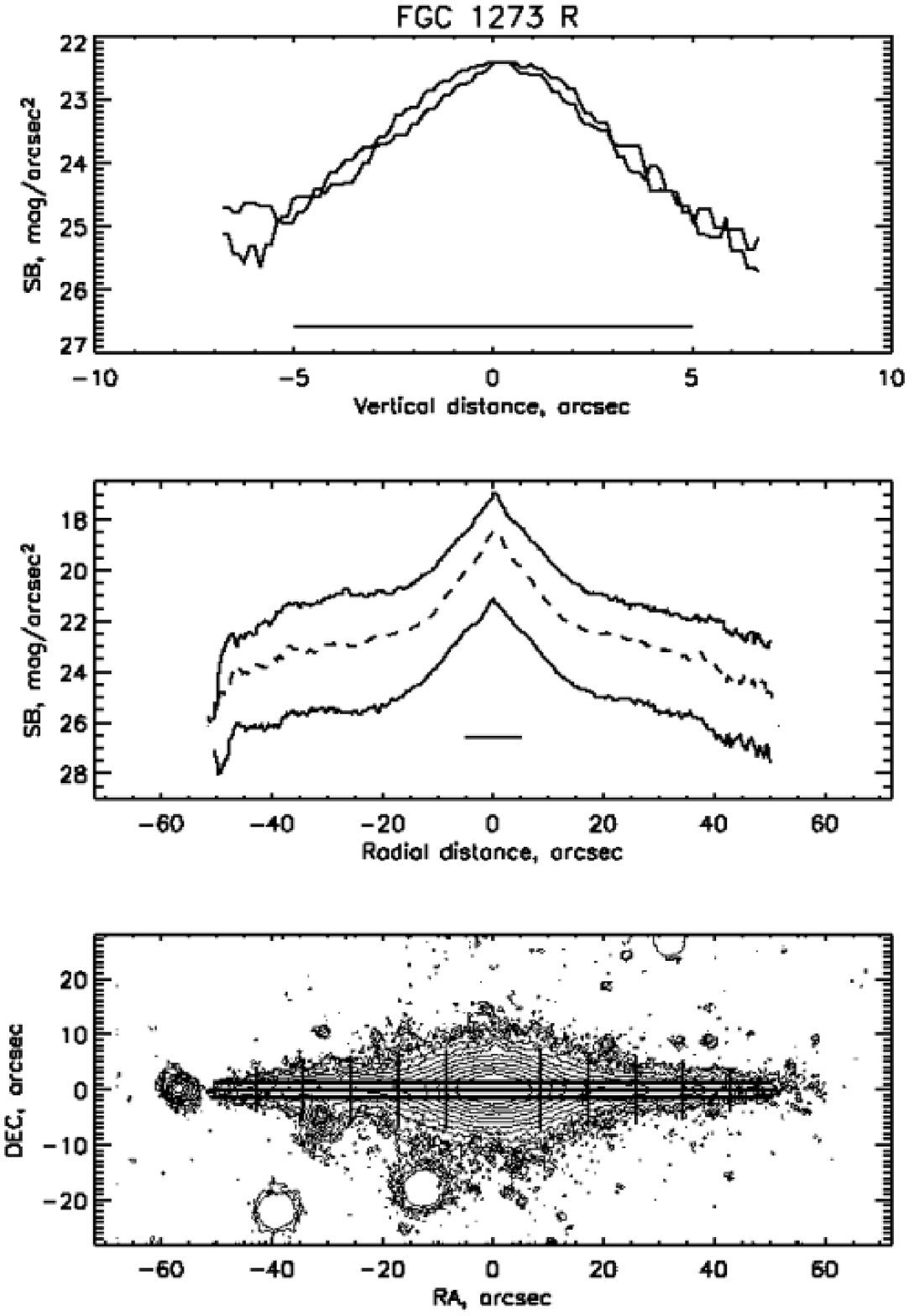}
\caption{
The same for FGC~1273. The isophotes correspond to    
20.5, 21., 21.5, 22., 22.5, 23.,23.5,24., 24.5,25., and 25.5
mag/arcsec$^2$.
}

\end{figure}


\clearpage
\begin{figure}
\epsscale{0.9}
\plotone{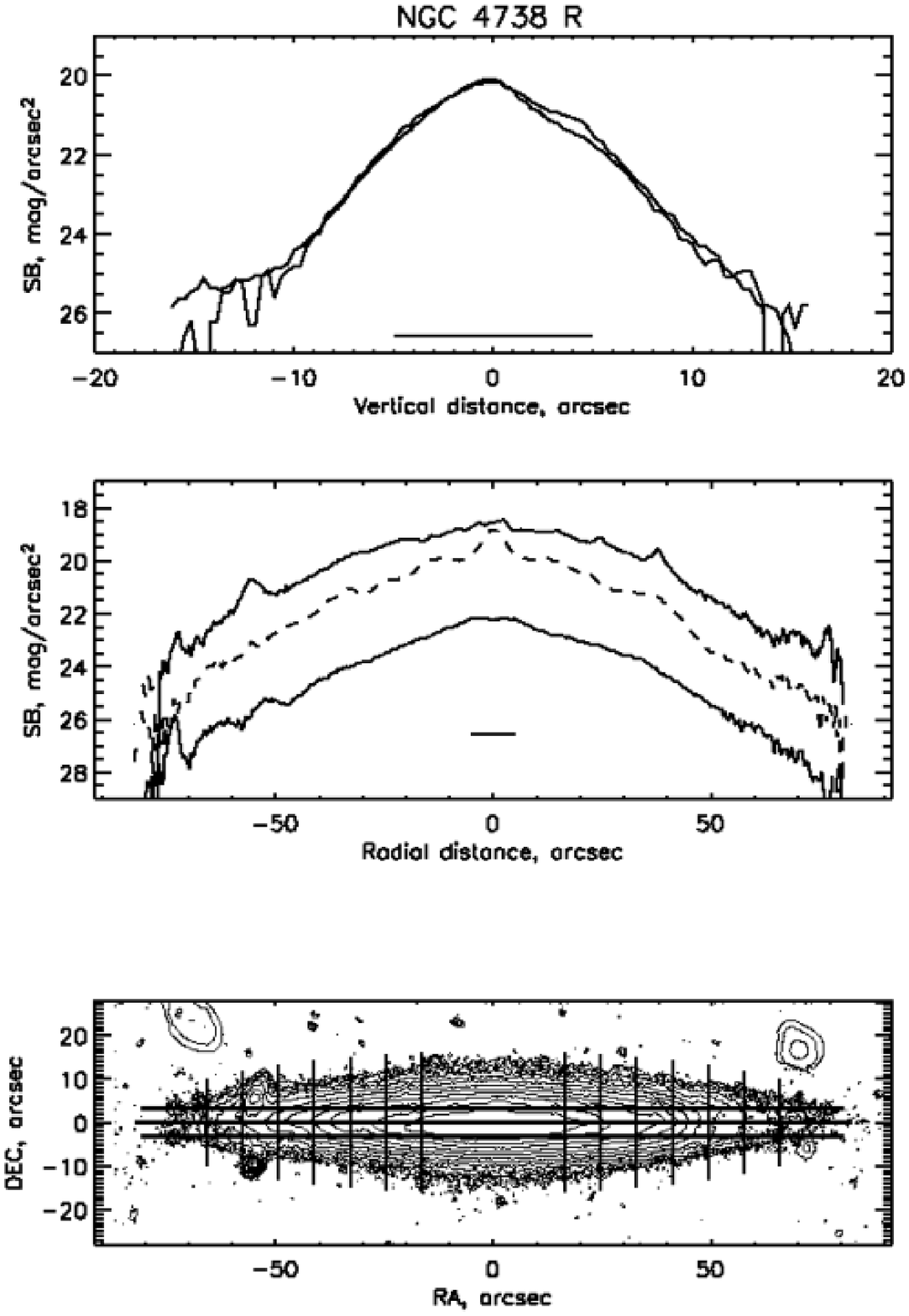}
\caption{
The same for NGC~4738. The isophotes correspond to    
20.5, 21, 21.5, 22, 22.5, 23, 23.5, 24, 24.5, 25, and 25.5
mag/arcsec$^2$.
\label{fig1k}
}

\end{figure}

\clearpage
\begin{figure}
\plotone{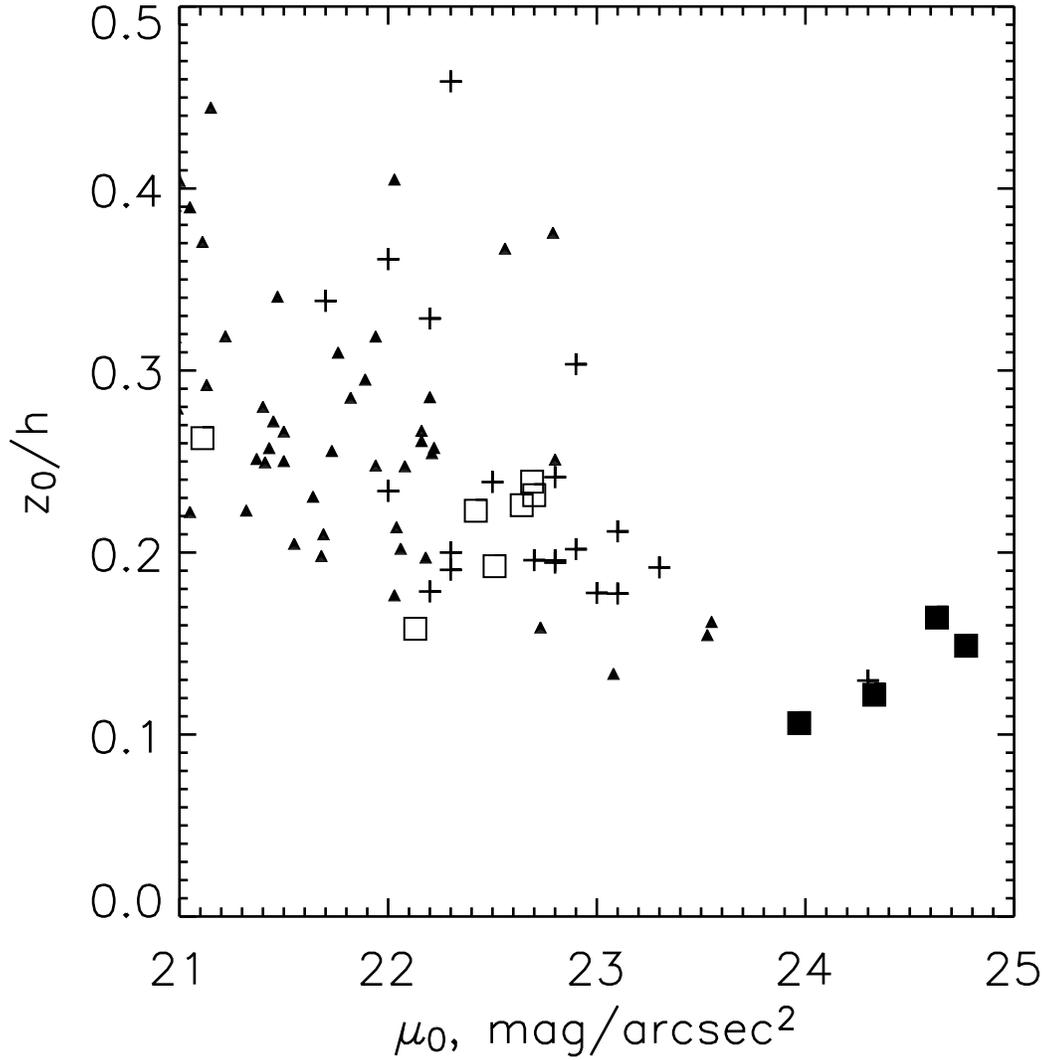}
\caption{
The vertical to radial scales ratio of the stellar
disks $z_0/h$ versus their central surface brightness 
$\mu_0$ in the R band. The objects from
our sample are denoted by squares. The open squares are for the HSB
subsample whereas the filled ones designate the LSB galaxies. 
The galaxies taken from \citet{B94} are shown as crosses, and the 
2MASS objects are denoted by the small filled triangles.
\label{fig2}}
\end{figure}

\clearpage
\begin{figure}
\plotone{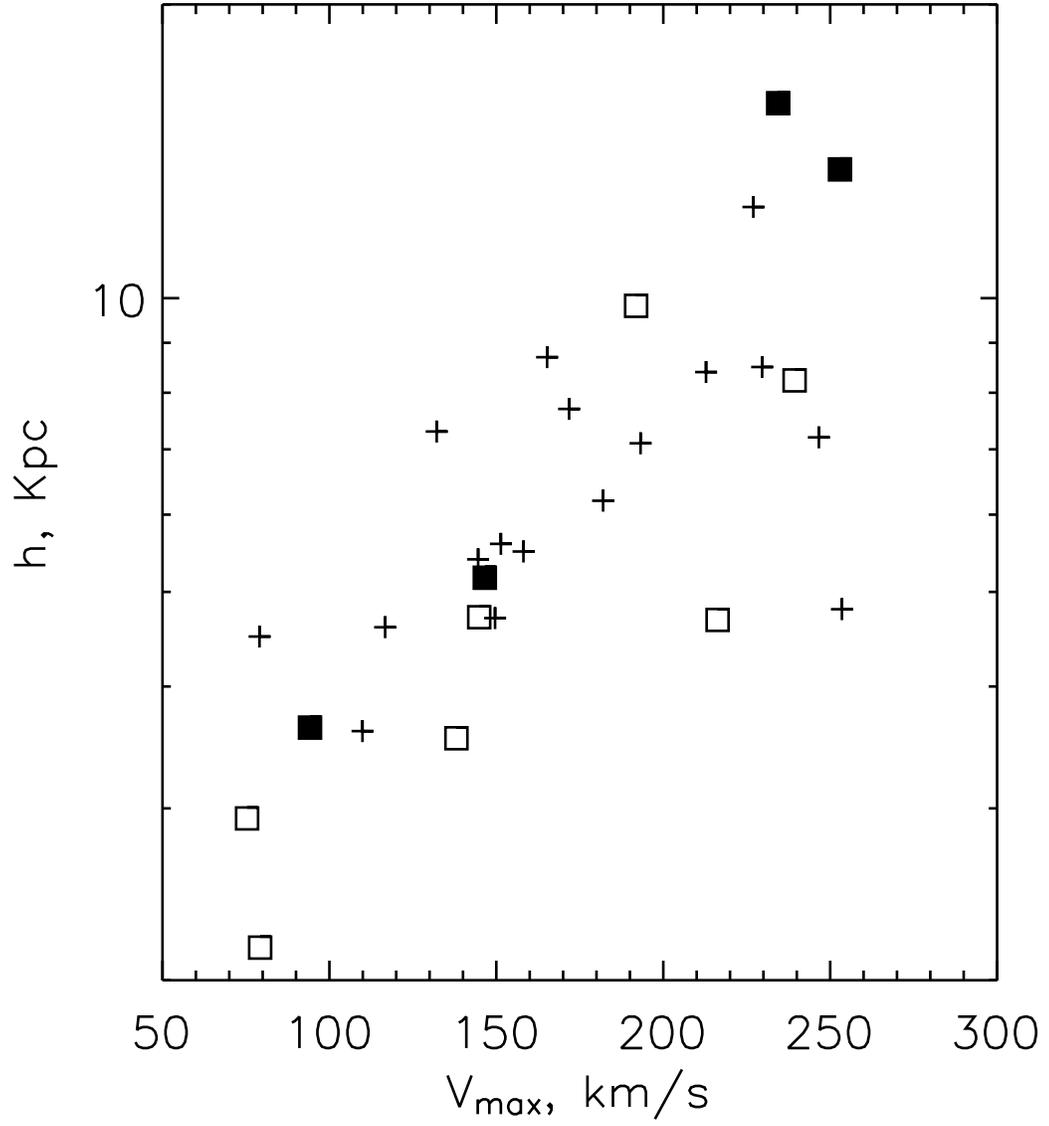}
\caption{
The radial scale length $h$ is well correlated with the maximum
rotational velocity $V_m$. The notation is the same as in Fig.\ref{fig2}.
\label{fig3}}
\end{figure}

\clearpage
\begin{figure}
\plotone{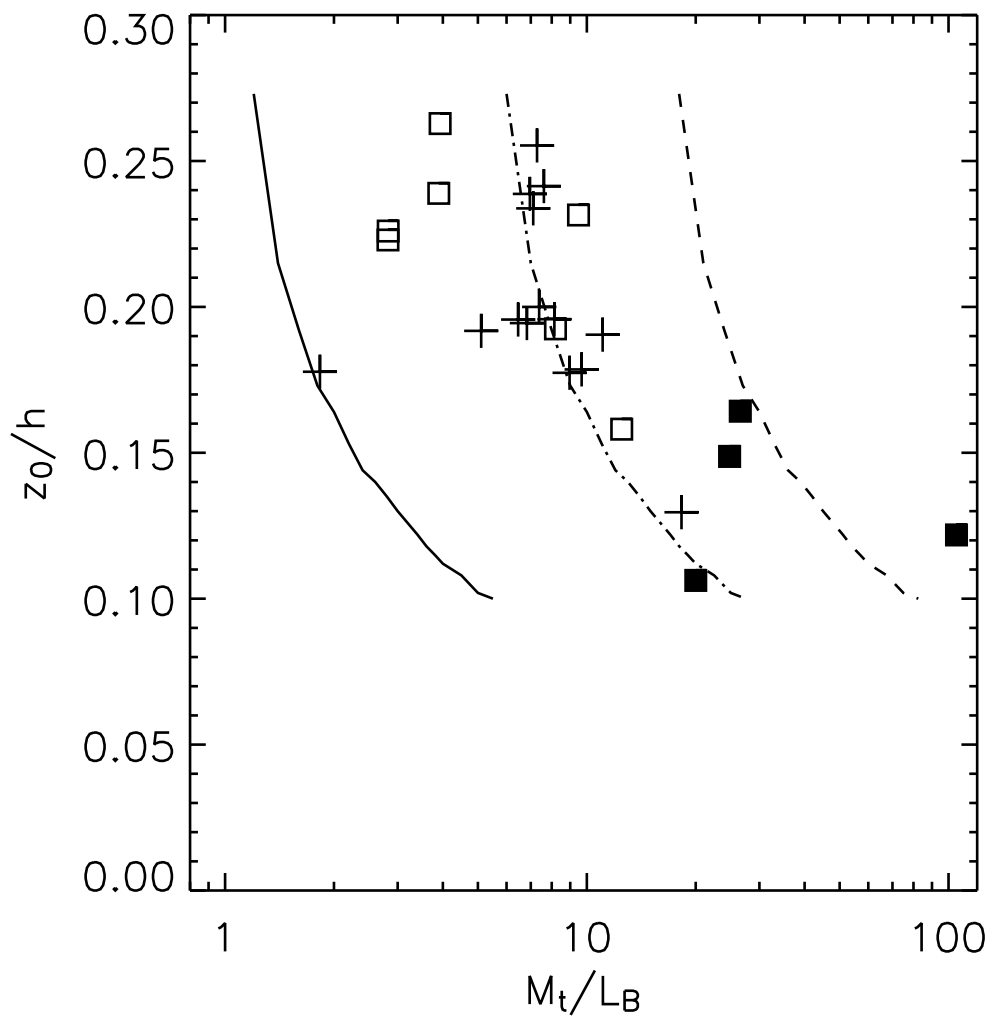}
\caption{
The ratio of the total mass to B-luminosity of disk $M_t/L_B$ 
(see in the text) plotted against the disk scales ratio $z_0/h$. 
The notation in the figure is the same as in Fig.\ref{fig2}. 
The three curves present the model values of $M_t/L_B$ which were 
calculated based on Fig.\ref{fig5} with the mass to light
ratios M/L of 1 (solid), 5 (dot-dashed) , and 15 (dashed).
\label{fig4}}
\end{figure}

\clearpage
\begin{figure}
\plotone{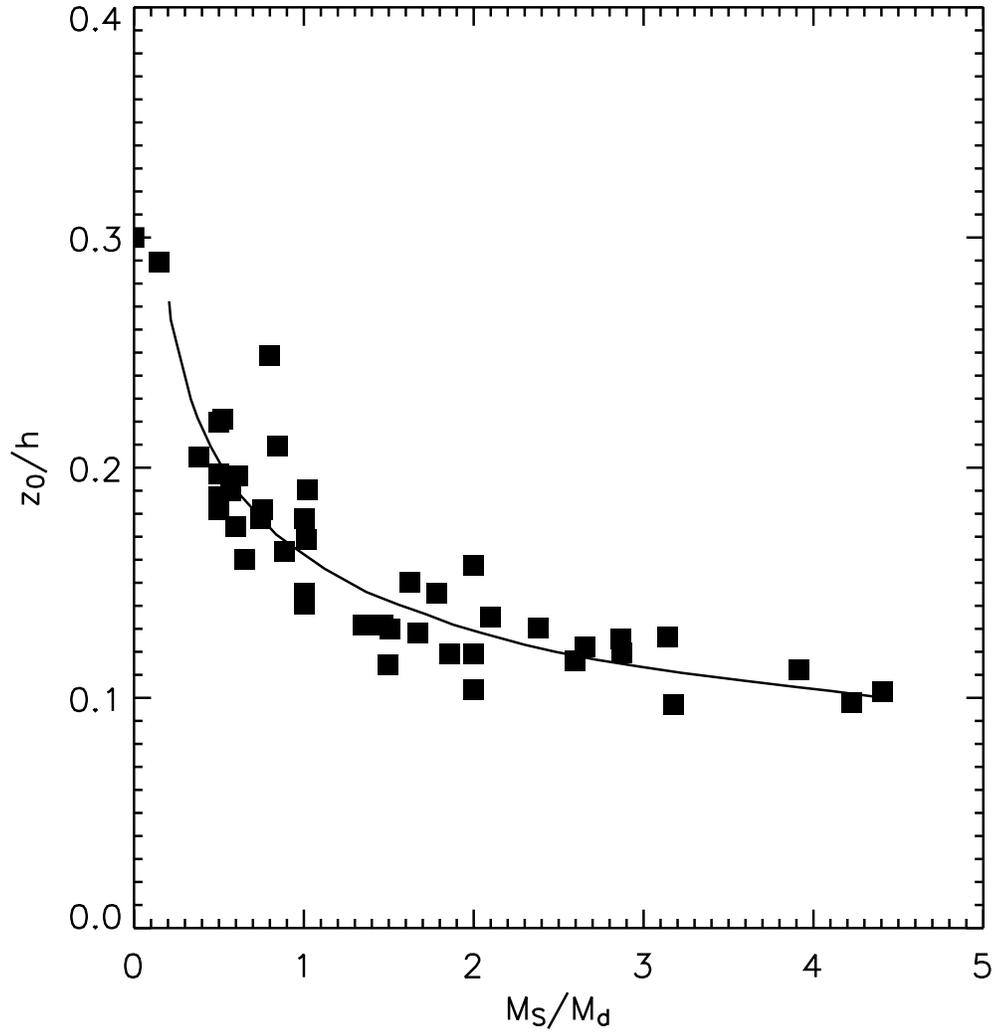}
\caption{
Relation between the stellar disk thickness $z_0/h$ and its relative
mass of the spherical component $M_s/M_d$ obtained from numerical
simulations (N body experiments). The figure is adopted from \citet{Khop01}.
\label{fig5}}
\end{figure}

\clearpage
\begin{figure}
\plotone{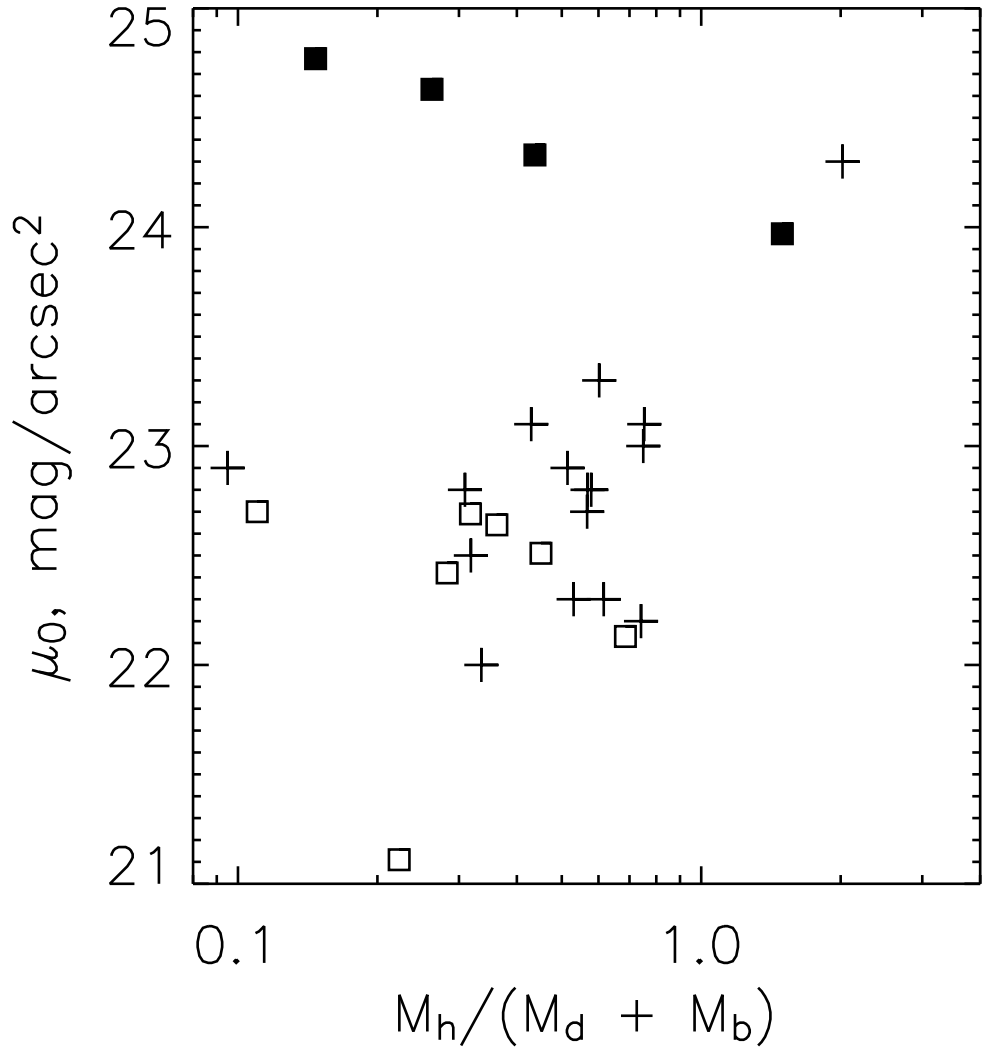}
\caption{
The ratio of dark-to-luminous mass $M_h/(M_d + M_b)$ for our galaxies.
The notation in the figure is the same as in Fig.\ref{fig2}.
\label{fig6}}
\end{figure}

\clearpage
\begin{figure}
\plotone{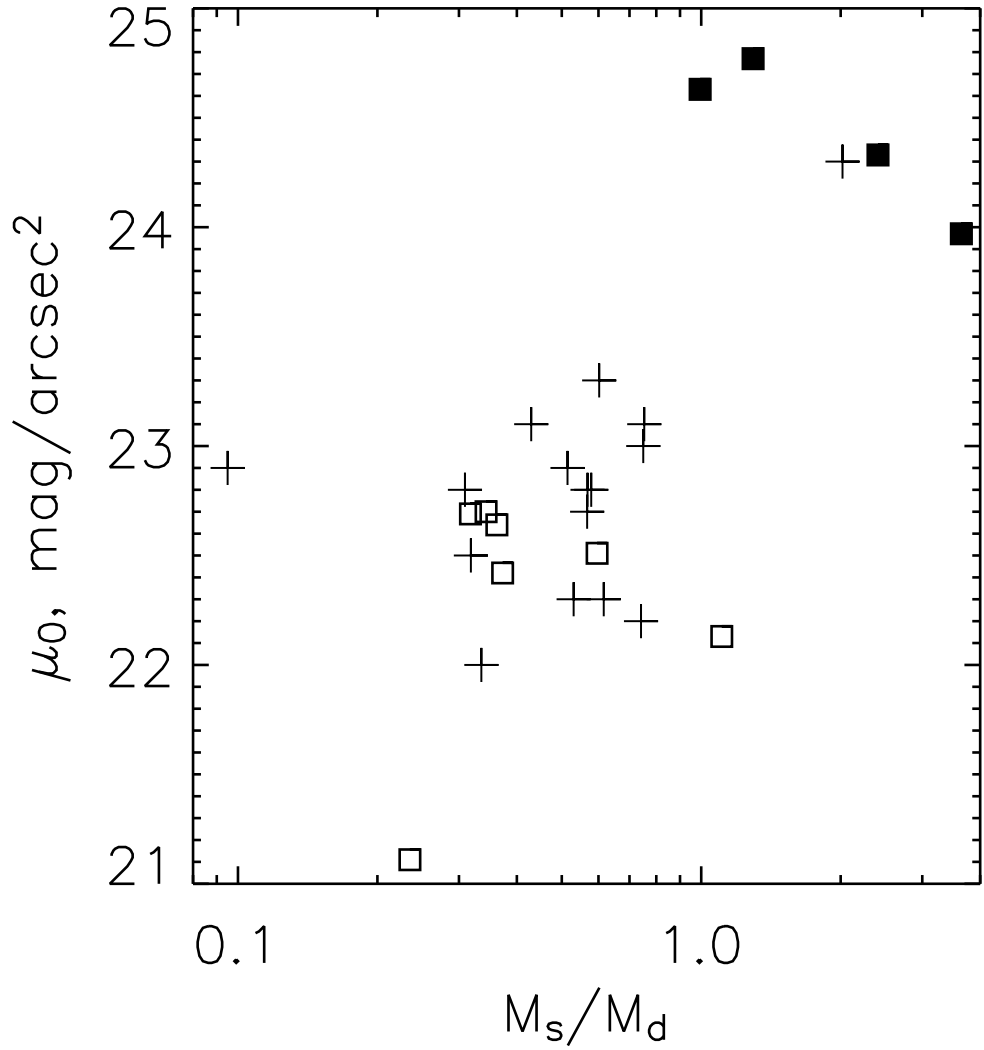}
\caption{
The spherical to disk mass ratio $M_s/M_d$ for our galaxies in 
dependence on the disk central surface brightness. We kept the 
same notation as in Fig.\ref{fig2}.
\label{fig7}}
\end{figure}


\begin{thebibliography}{}

\bibitem[Peletier \& Balcells (1996)]{peletier96}
Peletier, R., Balcells, M., 1996, \aj, 111, 2238

\bibitem[Barteldrees \& Dettmar(1994)]{B94}
Barteldrees, A., Dettmar, R.-J. 1994, \aaps, 103 ,475

\bibitem[Beijersbergen et al.(1999)]{beijersbergen99}
Beijersbergen, M., de Blok, W., and van der Hulst, J.
1999, \aa, 351, 903

\bibitem[Bizyaev(2000)]{B00}
Bizyaev, D. 2000, Sov. Astron. Lett., 26, 219

\bibitem[Bizyaev \& Mitronova(2002)]{BM02}
Bizyaev, D., Mitronova, S. 2002, \aap, 389, 795

\bibitem[de Blok et al.(1995)]{deBlok95}
de Blok, W., van der Hulst, J., Bothun, G. 1995, \mnras, 274, 235

\bibitem[de Blok et al.(1996)]{deBlok96}
de Blok, E., van der Hulst, T., \& McGaugh, S. 1996, AAS, 189, 8402

\bibitem[de Blok et al.(2003)]{deBlok03}
de Blok, W., Bosma, A., \& McGaugh, S. 2003, \mnras, 340, 657

\bibitem[Chung et al.(2002)]{Chung02}
Chung, A., van Gorkom, J., O'Neil, K., Bothun, G. 2002 \aj, 123, 2387

\bibitem[de Grijs \& van der Kruit(1996)]{deG96}
de Grijs, R., van der Kruit, P. 1996 \aaps, 117, 19

\bibitem[de Grijs et al.(1997)]{deG97}
de Grijs, R., Peletier, R., van der Kruit, P. 1997, \aap, 327, 966

\bibitem[de Jong(1996)]{deJ96} de Jong, R. 1996 \aap, 313, 377

\bibitem[Gadotti \& dos Anjos (2001)]{gadotti01}
Gadotti, D., dos Anjos, S. 2001 \aj, 122, 1298

\bibitem[Freeman (1970)]{freeman70}
Freeman, K. 1970, \apj, 160, 811

\bibitem[Graham(2001)]{Graham01} Graham, A. 2001, \mnras, 326, 543

\bibitem[Graham(2002)]{Graham02} Graham, A. 2002, \mnras, 334, 721

\bibitem[Holley-Bockelmann \& Mihos(2001)]{hm01}
Holley-Bockelmann, J., Mihos, J. 2001, AAS, 198, 08.15

\bibitem[Karachentsev et al.(1992)]{IDK92}
Karachentsev, I., Georgiev, Ts., Kajsin, S., Kopylov, A., Ryadchenko, V.,
Shergin, V. 1992, Astron \& Astrophys. Transact., 2, 265

\bibitem[Karachentsev et al.(1999)]{RFGC}
Karachentsev, I., Karachentseva, V., Kudrya, Y., et al. 1999, Bull. of
Special Astrophys. Obs., 47, 5

\bibitem[Kregel et al.(2002)]{Kregel02}
Kregel, M., van der Kruit, P., de Grijs, R. 2002, \mnras, 334, 646

\bibitem[Landolt(1992)]{Landolt92}
Landolt, A. 1992 \aj, 104, 340

\bibitem[Matthews(2000)]{Matt00}
Matthews, L. 2000, \aj, 120, 1764

\bibitem[Matthews \& Wood (2001)]{Matt01}
Matthews, L., Wood, K. 2001, \apj, 

\bibitem[McGaugh (1994)]{mcgaugh94}
McGaugh, S. 1994, \apj, 426, 135

\bibitem[McGaugh et al.(1995)]{mcgaugh95}
McGaugh, S., Schombert, J., and Bothun, G. 1995, \aj, 109, 2019

\bibitem[McGaugh et al.(2001)]{McG01} 
McGaugh, S., Rubin, V., \& de Blok, W. 2001, \aj, 122, 2381

\bibitem[Mikhailova et al.(2001)]{Khop01}
Mikhailova, E., Khoperskov, A., Sharpak, S. 2001, Conf. proc. "Stellar
Dynamics: From Classic to Modern", ed. Ossipkov \& Nikiforov, p. 147.

\bibitem[Nilson (1973)]{UGC}
Uppsala General Catalogue of Galaxies, 1973, Acta Universitatis Upsalienis,
Nova Regiae Societatis Upsaliensis.

\bibitem[Pfenniger et al. (1994)]{pfenniger94}  
Pfenniger, D., Combes, F., Martinet, L. 1994, \aap, 285, 79

\bibitem[Pohlen et al.(2002)]{pohlen02}
Pohlen, M., Dettmar, R.-J., Lutticke, R., and Aronica, G.
2002, \aap, 392, 807

\bibitem[Reshetnikov et al.(2003)]{Resh03}
Reshetnikov, V., Dettmar, R.-J., \& Combes, F. 2003 \aap, 399, 879

\bibitem[Romanishin et al.(1982)]{romanishin82}
Romanishin, W., Krumm, N., Salpeter, E., et al. 1982
\apj, 263, 94

\bibitem[Sprayberry et al.(1995)]{Sprayberry95}
Sprayberry, D., Bernstein, G., Impey, \& C., Bothun, G. 1995,
\apj, 438, 72

\bibitem[Tully et al (1998)]{tully98}   
Tully, R., Pierce, M., Huang, J.-S., et al. 1998,
\apj, 115, 2264

\bibitem[van der Kruit \& Searle(1981)]{vdK81} 
van der Kruit, P., Searle, L., 1981, \aap, 95, 105

\bibitem[van der Kruit et al.(2001)]{vdK01}
van der Kruit, P., Jimenez-Vicente, J., Kregel, M., \& Freeman, K.
2001, \aap, 379, 374

\bibitem[Xilouris et al.(1998)]{n891}
Xilouris, E., Alton, P., Davies, J., et al. 1998, \aap, 331, 894

\bibitem[Xilouris et al.(1999)]{Xilouris99}
Xilouris, E., Byun, Y., Kylafis, N., Paleologou, E., Papamastorakis,J.
1999, \aap, 344, 868

\bibitem[Zasov et al.(1991)]{Z91}
Zasov A., Makarov D., Mikhailova E. 1991, Sov. Astron. Lett., 17, 374

\bibitem[Zasov et al.(2002)]{Z02}
Zasov, A., Bizyaev, D., Makarov, D., Tyurina, N. 2002, Sov. Astron. Lett., 
28, 527

\bibitem[Zwaan et al.(1995)]{Zwaan95}
Zwaan, M., van der Hulst, J. , de Blok, W., \& McGaugh, S. 1995, 
\mnras, 273, L35

\end{thebibliography}
\end{document}